\documentclass[aps,prresearch,twocolumn,superscriptaddress,longbibliography]{revtex4-2}
\pdfoutput=1
\usepackage[utf8]{inputenc}
\usepackage[english]{babel}
\usepackage[T1]{fontenc}
\usepackage{amsfonts,amssymb, amsmath}
\usepackage{mathtools}
\usepackage{dsfont}
\usepackage{bm}
\usepackage{float}
\usepackage{physics}
\usepackage{graphicx}
\usepackage{xcolor}
\usepackage{relsize}
\usepackage{comment}
\usepackage{hyperref}

\begin{document}
\title{Quantum power iterations via tensor networks}

\author{V.~Akshay}
\affiliation{Terra Quantum AG, Kornhausstrasse 25, 9000 St. Gallen, Switzerland}

\author{A.~Berezutskii}
\email{berezutskii.aleksandr@gmail.com}
\affiliation{Terra Quantum AG, Kornhausstrasse 25, 9000 St. Gallen, Switzerland}

\author{Ar.~Melnikov}
\affiliation{Terra Quantum AG, Kornhausstrasse 25, 9000 St. Gallen, Switzerland}

\author{A.~Termanova}
\affiliation{Terra Quantum AG, Kornhausstrasse 25, 9000 St. Gallen, Switzerland}

\author{M.~R.~Perelshtein}
\email{mpe@terraquantum.swiss}
\affiliation{Terra Quantum AG, Kornhausstrasse 25, 9000 St. Gallen, Switzerland}

\begin{abstract}
In optimization, one of the well-known classical algorithms is power iterations. Simply stated, the algorithm recovers the dominant eigenvector of some diagonalizable matrix. Since numerous optimization problems can be formulated as an eigenvalue/eigenvector search, this algorithm features wide applicability. Operationally, power iterations consist of performing repeated matrix-to-vector multiplications (or \texttt{MatVec}) followed by a renormalization step in order to converge to the dominant eigenvalue/eigenvector. However, classical tensor network realizations can, in the worst case, exhibit an exponential growth of the tensor ranks with the number of iterations. In this study, we propose a quantum realization that offers a possible route around this bottleneck. Our method involves casting low-rank representations --- Matrix Product Operators (MPO) for matrices and Matrix Product States (MPS) for vectors --- into quantum circuits. Specifically, we recover a unitary approximation by variationally minimizing the Frobenius distance between a target MPO and an MPO ansatz wherein the tensor cores are constrained to unitaries. Such unitary MPO can be implemented as a quantum circuit with the addition of ancillary qubits. Thereafter, with appropriate initialization and post-selection on the ancillary space, we realize a single iteration of the classical algorithm. With our proposed methodology, power iterations can be realized entirely on a quantum computer via repeated, static circuit blocks. Moreover, by exploiting Riemannian optimization and cross-approximation techniques, our methodology becomes instance agnostic and thus allows one to address black-box optimization of functions admitting low-rank tensor network representations within the framework of quantum computing. We emphasize that the present work establishes the construction and analyzes the conditions under which it operates reliably; it does not demonstrate an end-to-end computational advantage over classical tensor network optimization, which we discuss explicitly.
\end{abstract}

\maketitle

\section{Introduction}

Modern approaches to quantum computation are centered on hybrid architectures, in which a classical co-processor and a noisy intermediate-scale quantum device (or NISQ) work in tandem to realize a computational objective~\cite{cerezo2021variational,mcclean2016theory,bharti2022noisy,biamonte2017quantum,preskill2018quantum, perelshtein2020large, perelshtein2023nisq, white_paper_tq}.
The advantage of using such architectures is two-fold: 
(1) it is expected that such setups may alleviate some of the systematic limitations of NISQ; and
(2) such models of computation may admit universality without the need for stringent error correction~\cite{lloyd2018quantum,morales2020universality,zhao2023universal}. 

However, even with these prospects, further circuit simplifications and/or compilations are necessary to overcome some of the hardware and classical processing limitations of hybrid algorithms~\cite{shaydulin2019hybrid, he2017decompositions, low2019hamiltonian,childs2012hamiltonian}. Moreover, since such approaches mostly constitute shallow quantum circuits, featuring low entanglement, these circuits can be efficiently emulated using tensor networks~\cite{orus2014practical,cirac2021matrix,eisert2008area,berezutskii2025tensor}. In fact, recent works demonstrate that tensor network emulations of quantum algorithms can be more effective than an actual quantum realization~\cite{gray2021hyper, liu2021closing, soley2021iterative, dolgov2020approximation, novikov2015tensorizing, berezutskii2025simulating}. These results question the possibility of attaining a quantum advantage via shallow-depth circuits~\cite{shor1999polynomial, harrow2009quantum, brassard2002quantum, plesch2011quantum}. 

Many of the recent advances in tensor networks feature promising algorithms based on low-rank representations. These techniques find applications in far-reaching areas, such as: in the study of ground states of many-body quantum systems~\cite{orus2014practical, cichocki2016tensor}, machine learning~\cite{huggins2019towards, abronin2024tqcompressor, naumov2023tetra, laskaris2023comparison}, and optimization~\cite{morozov2023protein, belokonev2023optimization, sagingalieva2023hybrid}. 
In tensor networks, one typically exploits the decomposition of tensors into a series of sub-tensors contracted by a set of virtual indices or bonds. By placing restrictions on the dimension of these virtual bonds (also known as sub-tensor ranks), one can recover low-rank approximations that are particularly suitable for the compression, representation, and processing of multidimensional data. Furthermore, since a quantum circuit can be viewed as a tensor network, these approaches can also be exploited for efficient simulation of quantum algorithms~\cite{perez2006matrix, glasser2019expressive, khoromskij2011d, gray2021hyper, ballani2014tree,pfeifer2014faster, zhang2022quantum,markov2008simulating,berezutskii2025simulating,berezutskii2025tensor}. The main contribution to the run-time complexity of tensor networks is the maximal sub-tensor ranks resulting from the prescribed decomposition. However, if such representations can be translated into quantum circuits, the bottleneck arising from large ranks can be alleviated~\cite{schon2005sequential,schon2007sequential, schwarz2012preparing,lubasch2020variational,perelshtein2020large}. It is precisely this regime that we explore. 

A well-known classical optimization algorithm, used in both applied and scientific areas, is power iteration. This method involves applying a standard discretization procedure to a function and then raising it to a large degree (with normalization) to retrieve the global maximum. However, when implemented using tensor networks, in the worst case, the ranks increase exponentially with the degree, limiting the scalability of the approach.
We address this issue by extending our previous results on encoding and preparing MPS on quantum computers~\cite{melnikov2023quantum, ran2020encoding} for the case of MPO~\cite{termanova2024tensor}. Specifically, we encode the MPO representation of an arbitrary, appropriately discretized function into a quantum circuit. Power iterations then become possible on a quantum computer by a simple concatenation of MPO circuit blocks and performing appropriate post-selection. Importantly, our technique circumvents the classical rank-growth bottleneck as \texttt{MatVec} operations are naturally carried out as physical processes on a general quantum computer~\cite{cleve2000introduction, koike2010time}. Provided that (i) the objective function admits an accurate low-rank MPS/MPO approximation, (ii) this approximation can be encoded into a unitary MPO with small Frobenius error, (iii) the post-selection probability remains acceptable (see Appendix~\ref{a1}), and (iv) the \emph{intermediate} iterates do not themselves admit accurate low-rank approximations, the circuit depth required for $p$ power iterations grows only polynomially in $n$, $p$ and $R$, whereas the ranks in a classical tensor network realization may grow exponentially with $p$. Condition (iv) is essential and is easily overlooked: it is \emph{not} satisfied by the smooth, low-dimensional benchmarks studied here, for which we verify in the Discussion that classical tensor train power iterations exhibit no rank blow-up at all. We accordingly claim no run-time advantage for the instances considered in this work, and we regard the identification of problem families satisfying (i)--(iv) simultaneously as the central open question raised by this construction.

We explore this methodology by considering three functions with increasing complexity for optimization:
1) the sine function, where we analytically recover the MPO representation and cast them into a quantum circuit; 2) the two-dimensional Ackley function, where we show optimization can still be addressed with erroneous and sub-optimal (specifically, rank-restricted) MPO representations; and finally 3) the two-dimensional Rosenbrock function, where we show that our proposed methodology is able to recover the complex optimization landscape of the function. Beyond these, we verify the approach in a higher-dimensional setting on the four-dimensional Ackley function discretized with 20 qubits. We emphasize that our proposed method is function-agnostic and therefore is also suitable for black-box optimization of objectives that admit accurate low-rank tensor network representations. Moreover, the circuit resources required per iteration grow only polynomially in the number of qubits at fixed ansatz rank and power. As we make precise in Sec.~\ref{qpm}, this does not by itself establish scalability: the post-selection probability and the number of measurements needed to identify the optimum are instance-dependent, and neither is polynomial in general.

\section{Background}
In this section, we briefly review the background needed to introduce quantum power iterations. In Subsection~\ref{b1}, we describe the classical algorithm using standard notation and emphasize the advantage of naturally realizing optimization, if one can implement the classical algorithm quantum mechanically. In Subsection~\ref{b2}, we introduce MPS and MPO, focusing on the efficiency arguments that come with such low-rank representations. We exploit these ideas in our quantum implementation; however, for a general review of these topics, we encourage the reader to review the following references:~\cite{golub2013matrix, orus2014practical, berezutskii2025tensor}.           

\subsection{Power Iterations}\label{b1}

The task of finding the largest eigenvalue of a diagonalizable matrix, after appropriate pre-processing, constitutes a general framework in which many optimization problems can be formulated~\cite{childs2021high, hou1978cubic, batsheva2023protes, arasu2002pagerank, panza2018application}. In order to employ this idea, we consider the following setting. 

Let $f(x)$ represent some real-valued function we wish to maximize over an input domain, $x \in [a,b]$.
Given access to $n$ qubits, one can exploit $2^n$ discretization points to represent $f(x)$ as a quantum state:
\begin{equation}\label{eq1}
    \ket{f} = \dfrac{1}{\sqrt{N}} \sum_{k = 1}^{2^n} f(x_{k}) \ket{k}, 
\end{equation}
where $x_k = \left( a + \left(k-1\right)\dfrac{b-a}{2^n}\right)$, 
$\left\{ \ket{k} \right\}_{k=1}^{2^n}$ are the computational basis for $n$ qubits and 
$N$ is the normalization constant that ensures  
$\lvert \braket{f}{f} \rvert = 1$. 

One can immediately see that the problem of maximization becomes
\begin{equation}\label{eq2}
    \max_{k} f(x_k) \equiv \max_{k} p_k,
\end{equation}
where $p_k = \lvert \braket{k}{f} \rvert^2$ are the measurement probabilities of state $\ket{f}$ with respect to projections onto the computational basis, and we assume $f\left(x_k\right) \geq 0$ so that the orderings of $f\left(x_k\right)$ and $p_k$ coincide (an assumption guaranteed by the pre-processing employed in this work). 
Therefore, if one could prepare such a state, measuring it would allow recovering a candidate optimum, up to errors caused by the discretization. Power iteration improves on this idea.

\noindent Consider a diagonal matrix,
\begin{equation}\label{eq3}
    \mathrm{H} = \sum_{k = 1}^{2^n} f(x_{k}) \ketbra{k}{k}.
\end{equation}
A standard power iteration gives: 
\begin{equation}\label{eq4}
    \ket{f^{\left(j+1\right)}} = \dfrac{\mathrm{H}^j \ket{f}}{\norm{ \mathrm{H}^j \ket{f}}},
\end{equation}
where $\norm{\cdot}$ is the 2-norm. 
Ideally, as $j \rightarrow \infty$, the resultant state becomes $\ket{k^{\ast}}$, where $k^{\ast} = \mathrm{arg} \: \max\limits_{k} \: p_{k}$ is the optimum~\cite{golub2013matrix, jung2001methode}. 
In order to practically realize this simple approach on a quantum computer, one must encode $\mathrm{H}$, as in Eq.~\eqref{eq3}, into a quantum circuit. However, this task is not straightforward. 

\subsection{Matrix Product States and Matrix Product Operators}\label{b2}

Low-rank tensor network representations are powerful tools used primarily to approximate and compress arbitrary tensors for efficient data processing~\cite{cirac2021matrix, oseledets2011tensor}. In the context of optimization, such low-rank representations are currently used in designing state-of-the-art, gradient-free, black-box optimizers~\cite{sozykin2022ttopt,soley2021iterative}. Of particular interest in this paper, is the fundamental advantage of compression brought about by a low-rank approximation. 
Specifically, an arbitrary state $\ket{\psi}$ of $n$ qubits is said to admit an MPS representation if it can be written as:
\begin{equation}\label{eq5}
    \resizebox{.89\hsize}{!}{$
    \displaystyle \ket{\psi}_{\mathrm{mps}} = \sum\limits_{\bm K \coloneqq k_1,k_2,\cdots ,k_n} \left[ {A^{ \left(1\right)}}_{k_1} {A^{ \left(2\right)}}_{k_2} \cdots {A^{ \left(n\right)}}_{k_n}\right] \ket{\bm K},
    $}
\end{equation}
where $ \bm K \coloneqq k_1,k_2,\cdots ,k_n \: \lvert \:  k_j \in \left\{ 0,1 \right\} ~\forall~ j \in [1,n]$, index the computational basis for $n$ qubits, and $\left\{ {A^{ \left(j\right)}}_{k_j} \right\}_{j=1}^{n}$ are some set of indexed matrices (also known as cores). Note that for Eq.~\eqref{eq5} to hold, the border cores $\left\{ {A^{ \left(1\right)}}_{k_1}, {A^{ \left(n\right)}}_{k_n} \right\} $ must have unit rank. 

An important parameter in low-rank representations is the maximal ranks defined as, $r \coloneqq \max\limits_{j} \: \mathrm{rank} \left( {A^{ \left(j \right)}}_{k_{j}} \right)$. It can be shown that any arbitrary state $\ket{\psi}$, admits an MPS representation, $\ket{\psi}_{\mathrm{mps}}$, albeit with maximal ranks scaling exponentially with the number of qubits \cite{cirac2021matrix, biamonte2017tensor}. However, this is for exact representations. When considering low-rank approximations, a wide class of states can be represented with $n$-independent or polynomially scaling MPS ranks~\cite{oseledets2013constructive}. A key advantage here is that such states can be represented with at most $2nr^2$ elements (see Eq.~\eqref{eq5}) instead of $2^n$~\cite{oseledets2011tensor}. Furthermore, computing quantum attributes such as the inner product can be done with polynomial run-times on a classical computer if the states admit rank-efficient MPS representations~\cite{oseledets2011tensor}. 

Similar to MPS, MPO are low-rank representations of linear operators. Let $\mathrm{H}$, be some linear operator on the $n$ qubits. We say $\mathrm{H}$ admits an MPO representation if it can be written as:
\begin{equation}\label{eq6}
    \resizebox{.89\hsize}{!}{$
        \displaystyle \mathrm{H}_{\mathrm{mpo}} = \sum\limits_{\bm K, \bm J} \left[ {A^{ \left(1\right)}}_{k_1}^{j_{1}} {A^{ \left(2\right)}}_{k_2}^{j_{2}} \cdots {A^{ \left(n\right)}}_{k_n}^{j_{n}}\right] \ketbra{\bm K}{\bm J},
        $}
\end{equation}
where $k_1,k_2,\cdots,k_n$ and $j_1,j_2,\cdots,j_n$ represent qubit indices and 
$\left\{ {A^{ \left(l\right)}}_{k_l}^{j_{l}} \right\}_{l=1}^{n}$ are a set of tensor cores.

All the efficiency arguments for the MPS follow through to MPO and, furthermore, one can use clever constructions to show that operations such as \texttt{MatVec}, calculating norms, the Hadamard product, etc. can be computed efficiently with these representations~\cite{oseledets2011tensor}. The standard procedure for constructing low-rank representations involves performing repeated truncated singular value decomposition (SVD) or by using cross-approximation methods~\cite{huber2017randomized,biamonte2017tensor,oseledets2010tt}. The latter allow one to construct approximate representations by sampling function values at a randomly chosen set of input indices, so that at no point in the MPS/MPO approximation is one forced to store or access all $2^n$ elements of the state; for the moderately sized instances studied in this paper, we construct the representations by truncated SVD, which is exact and convenient at these scales.

\section{Quantum Power Iterations}\label{qpm}

\begin{figure*}[!ht]
    \centering
    \includegraphics[width=\textwidth]{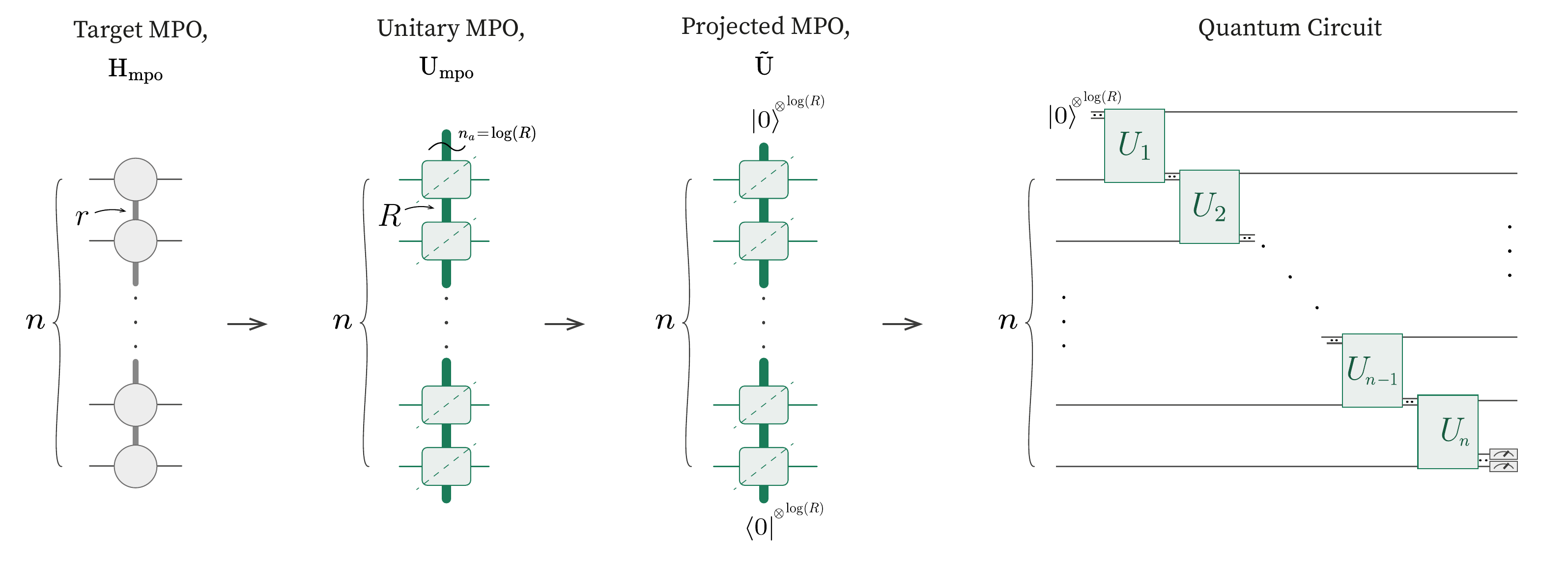}
    \caption{A schematic diagram for casting an arbitrary MPO into a quantum circuit. Left to Right: A target MPO is approximated by a unitary MPO of larger rank. Here, the dashed lines over MPO cores indicate the partitioning of indices with respect to which the cores are constrained to unitaries. Adhering to appropriate reshaping, the unitary MPO can be viewed as a quantum circuit with a step-like architecture.}
    \label{Fig1}
\end{figure*}

To motivate our approach, we briefly describe the pitfalls associated with a tensor network realization of power iterations. 

Let $\ket{f}_{\mathrm{mps}}$, correspond to a low-rank MPS representation for some function $f\left(x\right)$ as in Eq.~\eqref{eq1}. Here we use cross-approximation techniques to obtain the MPS cores as they only require polynomially many samples to adequately approximate $f\left(x\right)$ \cite{oseledets2010tt}. Note that for power iterations, one also requires a diagonal matrix $\mathrm{H}$, as in Eq.~\eqref{eq3}. An MPO representation, $\mathrm{H}_{\mathrm{mpo}}$, can easily be constructed from the MPS cores by enforcing a diagonal restriction: 
\begin{equation}\label{eq7}
    \resizebox{.89\hsize}{!}{$
    \displaystyle \mathrm{H}_{\mathrm{mpo}} = \sum\limits_{\bm K} \left[ {A^{ \left(1\right)}}_{k_1} {A^{ \left(2\right)}}_{k_2} \cdots {A^{ \left(n\right)}}_{k_n}\right] \ketbra{\bm K}{\bm K},
    $}
\end{equation}
Here, $\left\{ {A^{ \left(j\right)}}_{k_j} \right\}_{j=1}^{n}$ are obtained straightforwardly from $\ket{f}_{\mathrm{mps}}$. Therefore, this diagonal MPO must have the same ranks as the MPS.

To realize power iteration, one considers the action of $\mathrm{H}_{\mathrm{mpo}}$ on say, $\ket{f}_{\mathrm{mps}}$. It is straightforward to see that this action results in a state which can easily be represented as an MPS, however, with squared ranks. Therefore, subsequent iterations of this naive strategy will result in MPS ranks to scale exponentially, making the approach intractable. This is where quantum computation can be exploited. In a universal quantum computer, unitary matrices are compiled into sequences of quantum gates~\cite{williams1998explorations}. Applications of these gates to the initial quantum state (essentially, vector-matrix multiplication) is naturally realized irrespective of their ranks, with run-times scaling polynomially in the gate count~\cite{cleve2000introduction, koike2010time}. To exploit this notion, one must encode low-rank representations into quantum circuits.

In the case of MPS, encoding strategies involve an orthogonalization procedure to obtain MPS representations with isometric cores (also known as canonicalization) --- a standard construction in the tensor network literature~\cite{vidal2003efficient, perez2006matrix, schollwock2011density, cirac2021matrix} --- which was employed for quantum circuit encoding in Ref.~\cite{melnikov2023quantum}. These isometric cores can then be extended to obtain unitary matrices by the addition of ancillary spaces. Thereafter, a circuit realization becomes straightforward~\cite{lubasch2020variational, ran2020encoding, rudolph2023decomposition}. However, a direct generalization of this method for MPO is a known challenge~\cite{termanova2024tensor}. Here, we address this issue by employing a variational search.     

Let $\bm\Omega$ be the space of MPO in which each individual core is constrained to satisfy unitarity. We call this MPO ansatz a ``unitary MPO''. Given a target, $\mathrm{H}_{\mathrm{mpo}}$, of rank $r$, our aim is to approximate it with a unitary MPO of rank $R \geq r$. In circuit description, the virtual bonds of the unitary MPO can be thought of as ancillary spaces; a circuit realization then becomes possible with the addition of $\log_2\left(R\right)$ ancillary qubits in addition to $n$ system qubits. The variational search can now be formulated in the following way.
\begin{figure*}[!ht]
    \centering
    \includegraphics[width=\textwidth]{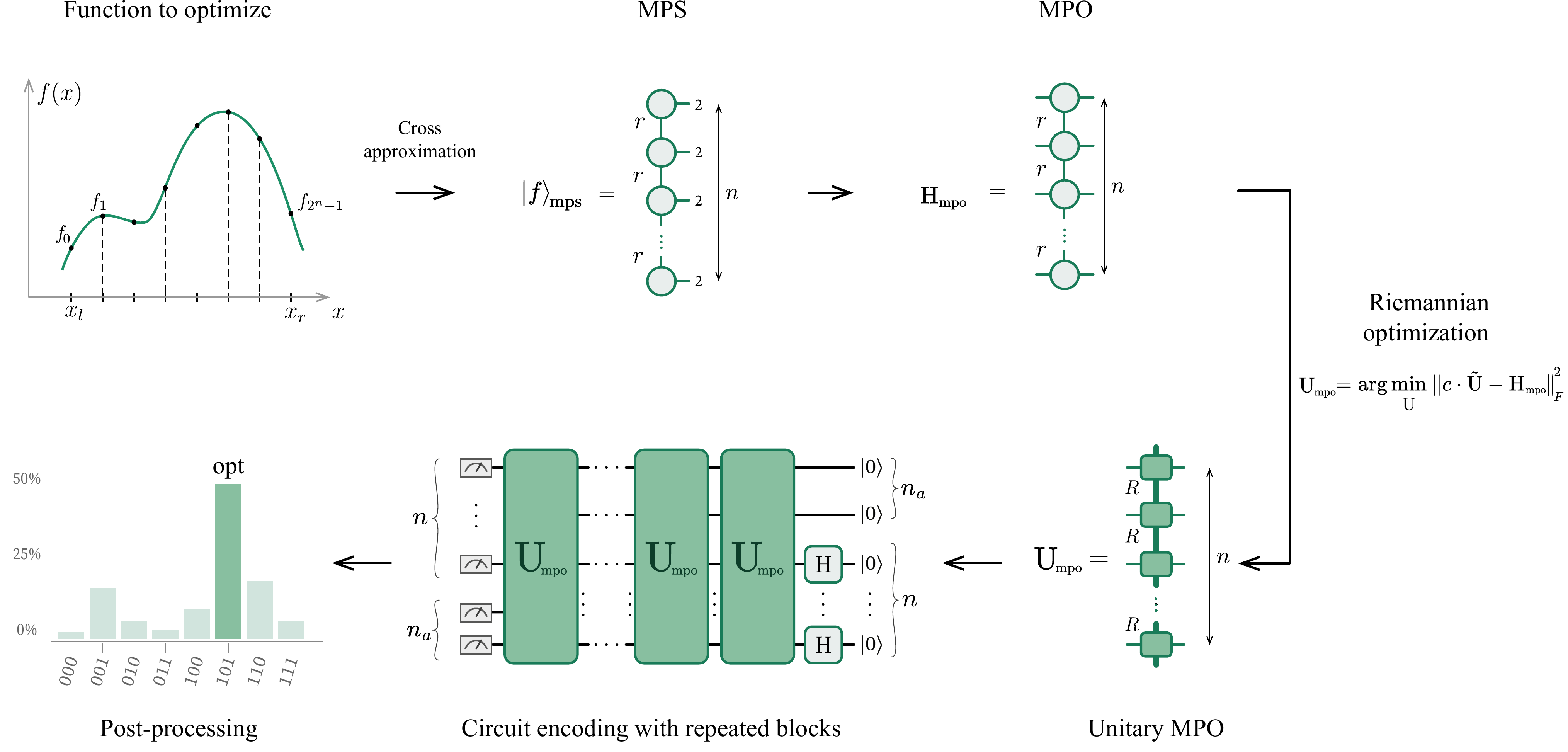}
    \caption{An illustration of quantum power iterations as a general purpose algorithm for optimization.}
    \label{Fig2}
\end{figure*}

Given a point $\mathrm{U} \in \bm \Omega$, we contract the qubit indices representing the ancillary space with its arbitrary choice of initialization to recover a linear operator, say $\tilde{\mathrm{U}}$, of the same dimension as $\mathrm{H}_{\mathrm{mpo}}$:
\small
\begin{equation}\label{eq8}
    \begin{split}
        \tilde{\mathrm{U}} &= \displaystyle \left(\mathds{1}^{\otimes n} \otimes \bra{0}^{ \otimes \log_2\left(R\right)} \right) \mathrm{U} \left(\mathds{1}^{\otimes n} \otimes \ket{0}^{\otimes \log_2\left(R\right)}  \right)\\
    \end{split}
\end{equation}
\normalsize
We then variationally minimize the Frobenius distance to the target as:   
\begin{equation}\label{eq9}
   \left( c^\ast, \mathrm{U}_{\mathrm{mpo}} \right) \in \arg \min_{\substack{c \in \mathbb{R} \\ \mathrm{U} \in \bm \Omega}} \norm{c \cdot \tilde{\mathrm{U}} - \mathrm{H}_{\mathrm{mpo}}}_{F}^{2}, 
\end{equation}
where $c^\ast \in \mathbb{R}$ is a constant scale factor. Minimizing Eq.~\eqref{eq9} and recovering an optimal $\mathrm{U}_{\mathrm{mpo}}$ is addressed using established Riemannian optimization techniques~\cite{luchnikov2021riemannian,kressner2016preconditioned}.  

The introduction of ancillary qubits is also justified by the fact that the matrix $\mathrm{H}_{\mathrm{mpo}}$ we are approximating, is not unitary, and for the implementation of non-unitary matrices on a quantum computer, ancillary qubits and post-selection is always necessary. In Fig.~\ref{Fig1}, we sketch out the circuit realization for $\mathrm{U}_{\mathrm{mpo}}$. Note that the circuit geometry features a step-like architecture with gates acting non-trivially on $\left(\log_2\left(R\right) + 1\right)$ qubits. By repeating this circuit block multiple times, one can easily realize power iterations on a general quantum computer. 

However, current quantum architectures cannot realize such arbitrary multi-qubit operations and therefore must be compiled into a sequence of single and two-qubit realizable operations~\cite{rakyta2022efficient}. For our particular case, it can be shown that the number of CNOT gates in the compiled circuit scale $ \sim n R^2$. Therefore our approach is still effective when considering low-rank class of MPO.

Another area of concern is the probabilistic nature of our approach due to the requirement of post-selection on the ancillary spaces. Though we show that the algorithmic success, in our considered setting, is independent of the number of qubits (see Appendix~\ref{a1}), amplitude amplification techniques may be needed for general cases \cite{brassard2002quantum}. 

In Fig.~\ref{Fig2}, we present a visual representation of our methodology for solving optimization problems through quantum power iterations. Algorithmically, the method can be summarized as follows: 

\begin{enumerate} 
    \item Prepare the initial state: $ \ket{+}^{\otimes n} \otimes \ket{0}^{\otimes n_a}$, where $n_a = \log_2\left(R\right)$. Here we skip preparing $\ket{f}_{\mathrm{mps}}$, as it is easier to prepare $\ket{+}^{\otimes n}$.
    \item Apply the circuit, $ \mathrm{U}_{\mathrm{mpo}} = \left(\prod_{k = 1}^{n}\limits U_{k}\right)$.
    \item Sequentially append the circuit with new ancillary spaces initialized as: $\ket{0}^{\otimes n_a}$, until the required power $p$ is attained. 
    \item Measure all ancillary spaces in the computational basis.
    \item If the measurement outcomes correspond to the bitstring: $ 0^{\times \left(p\cdot n_a \right)}$, measure the system qubits and recover bit-strings corresponding to a candidate optimum. In case of an unsuccessful outcome, return to Step 1.
    \item Repeat Steps 1-5 to collect measurement statistics and infer the candidate optimum as the most likely outcome (see Appendix~\ref{a2}).  
\end{enumerate}

Finally, we also provide estimates (see Table~\ref{tab:1}) for the run-times associated with each step in our proposed method. These ingredients assemble into the following end-to-end cost estimate. The classical pre-processing comprises: (i) the construction of the MPS approximation, requiring $\mathcal{O}\left(nr^2\right)$ function evaluations via cross-approximation; (ii) the variational search for the unitary MPO, requiring $N_{\mathrm{res}} \cdot N_{\mathrm{it}}$ gradient steps of cost $\mathcal{O}\left(nR^3\right)$ each, where $N_{\mathrm{res}}$ is the number of random restarts and $N_{\mathrm{it}}$ the number of iterations per restart (in our experiments, $N_{\mathrm{res}} \leq 3$ and $N_{\mathrm{it}} = 10^4$ sufficed whenever the reliability criterion discussed in the Results section was satisfiable); and (iii) the gate compilation, performed once per problem. On the quantum side, a single execution of $p$ power iterations has depth $\mathcal{O}\left(npR^2\right)$ and post-selects successfully with probability $P_{\mathrm{success}}$ (Appendix~\ref{a1}), while identifying the discretized optimum from the measurement statistics requires a number of successful outcomes $M$ that scales inversely with the squared gap (Appendix~\ref{a2}). The expected total quantum run-time is therefore $\mathcal{O}\left(npR^2 \cdot M / P_{\mathrm{success}}\right)$. We stress that only the explicit circuit-resource factor $npR^2$ is polynomial in $n$ at fixed $R$ and $p$, and that this alone does not make the method scalable: the ansatz rank $R$ required to reach a prescribed fitting accuracy, the number of gradient steps $N_{\mathrm{it}}$, the number of successful outcomes $M$ and the post-selection probability $P_{\mathrm{success}}$ are all instance-dependent and are not controlled by our analysis. Two of them are in fact provably \emph{not} polynomial in general: $P_{\mathrm{success}}$ decays exponentially in $p$ (Appendix~\ref{a1}), and the number of measurements needed to identify the discretized optimum \emph{exactly} grows exponentially in $n$ for a smooth objective, since refining the grid makes the two largest values increasingly degenerate (Appendix~\ref{a2}). The latter is a statement about exact identification of the argmax rather than about optimization: the probability of returning a grid point within a fixed tolerance of the continuum optimizer is independent of $n$ (Appendix~\ref{a2}). The function-dependent factors $M$ and $P_{\mathrm{success}}$ are quantified for our benchmarks in Table~\ref{tab:2}. As one may generally infer, our proposed methodology is adaptable to arbitrary functions that satisfy the low-rank requirements discussed above. Therefore, we believe our results constitute a step towards black-box optimization in quantum computing for this function class, and we expect further advances as quantum technologies continue to evolve.

\begin{table}[h]
    \centering
    \begin{tabular}{p{0.615\linewidth} c}
        \textbf{Description} & \textbf{Run-times} \\
        \hline
        Discretizing $f(x)$ and recovering a rank $r$ MPS representation via cross approximation & $\mathcal{O}\left(nr^2\right)$ \\
        Obtaining a diagonal MPO from an MPS representation & $\mathcal{O}\left(nr^2\right)$ \\
        A single evaluation of the cost function with a unitary MPO ansatz of rank $R$ (see Eq.~\eqref{eq9}) & $\mathcal{O}\left(nR^3\right)$ \\
        Compiled CNOT gate counts needed for realizing unitary MPO & $\mathcal{O}\left(nR^2\right)$ \\
        Circuit depth needed for $p$ quantum power iterations & $\mathcal{O}\left(npR^2\right)$ \\ 
        Total run-time in case of a successful outcome & $\mathcal{O}\left(npR^2\right)$ \\ \\
    \end{tabular}
    \caption{Run-time estimates for quantum power iteration.}
    \label{tab:1}
\end{table}

\section{Results}

We consider three functions with increasingly complex landscapes in order to test our proposed method for optimization, and subsequently extend the study to a four-dimensional benchmark. As a preliminary example, we first consider the $\sin\left(x\right)$ function over the domain $x \in [ 0, \pi ]$. 

Next, we consider the two-dimensional Ackley function defined as: 
\begin{equation}\label{eq10}
    \begin{split}
        f\left(x,y\right) &= -20\exp\left(-0.2\sqrt{0.5\left( x^2 + y^2 \right)}\right) \\
        &- \exp\left( 0.5\left( \cos 2\pi x +  \cos 2\pi y \right) \right) \\
        &+ 20 + e,
    \end{split}
\end{equation}
over the domain, $x,y \in [-5,5] $. 
Here, the function attains a global minimum $f\left( 0,0 \right) = 0$.
Finding this global minimum is a well-known non-convex optimization problem which poses difficulty for gradient based optimizers~\cite{back1996evolutionary}. 

Finally, we consider the two-dimensional Rosenbrock function defined as: 
\begin{equation}\label{eq11}
    f(x,y) = \left(1-x\right)^2 + 100\left(y - x^2\right)^2 
\end{equation}
over the domain $x,y \in [ -2.5, 2.5]$. 
The function attains a global minimum at $[1,1]$ along a narrow banana shaped valley.

\begin{figure*}[ht!]
    \centering
    \includegraphics[width = 0.49\textwidth]{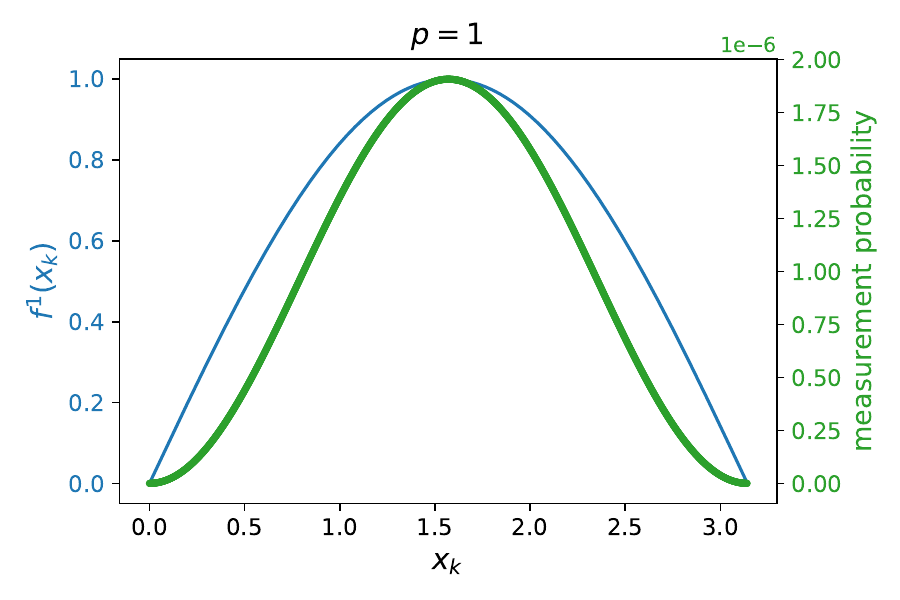}
    \includegraphics[width = 0.49\textwidth]{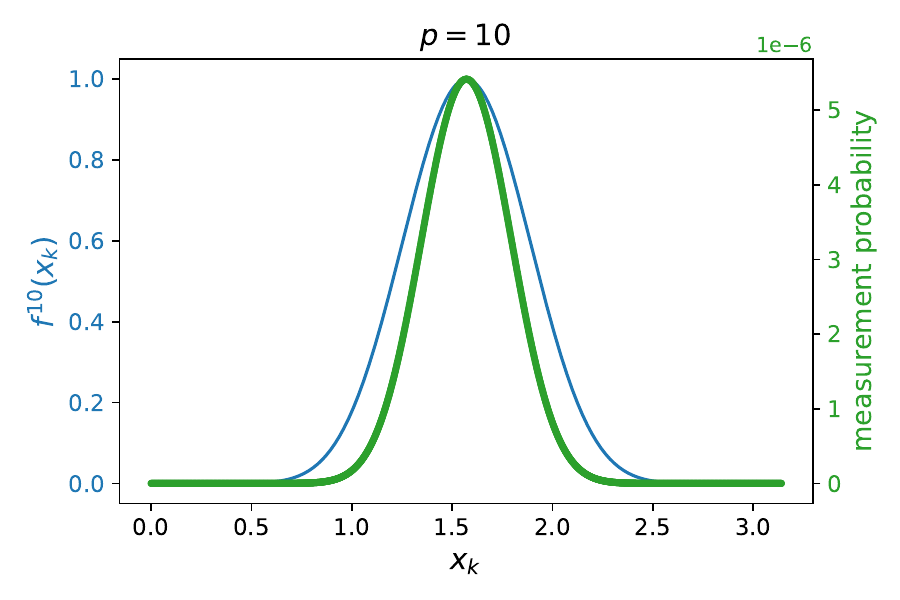}\\
    \includegraphics[width = 0.49\textwidth]{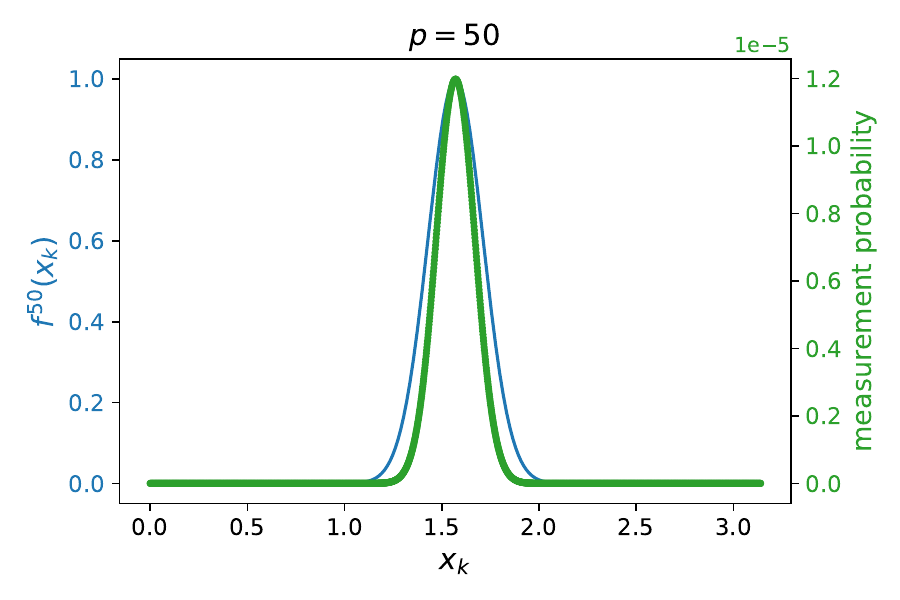}
    \includegraphics[width = 0.49\textwidth]{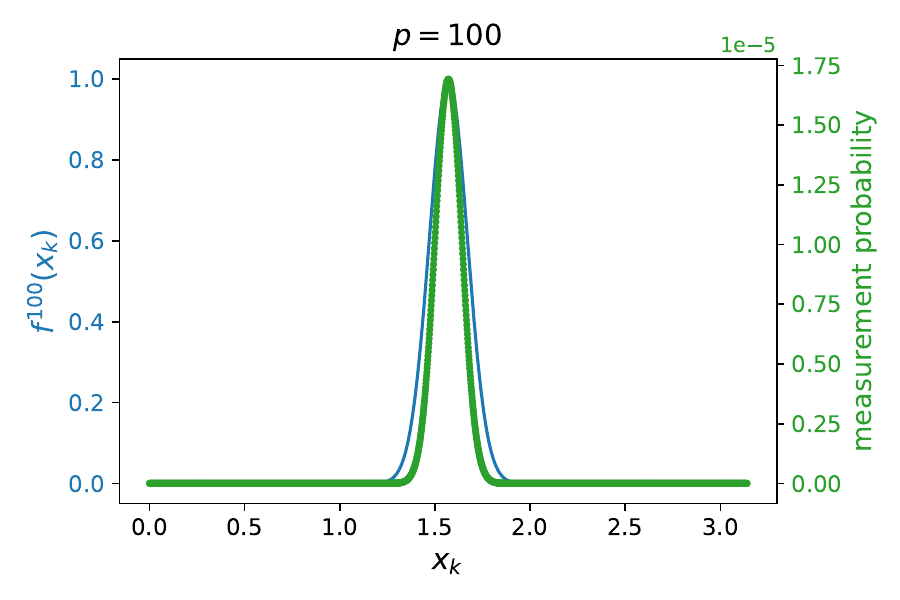}
    \caption{Twin axes plots representing the quantum power method. 
    In each subplot, the left $y$-axis corresponds to the exact values for the powers of sine (blue curve) and on the right $y$-axis, we indicate the measurement probabilities (green dots) obtained by simulating quantum power iterations.}
    \label{Fig3}
\end{figure*}

The three functions considered above were chosen to illustrate three different scenarios for testing the performance of our approach. For the sine function, which we directly maximize, no pre-processing is required. The Ackley and Rosenbrock benchmarks are minimization problems and are converted into maximization problems via the transformation
\begin{equation}\label{eq:preproc}
    \tilde{f}\left(x,y\right) = 1 - \tanh\left( \dfrac{f\left(x,y\right)}{\kappa} \right), \quad \kappa = f\left(a,a\right),
\end{equation}
where $\left(a,a\right)$ is the corner of the optimization domain. Since $\tanh$ is strictly monotone and both benchmarks attain the global minimum $f = 0$, Eq.~\eqref{eq:preproc} maps the location of the global minimum of $f$ to the global maximum $\tilde{f} = 1$ while preserving the ordering of all function values, and yields a bounded objective $\tilde{f} \in \left(0, 1\right]$. We note that the choice of the pre-processing directly affects the performance of the method: the scale $\kappa$ controls the contrast between the optimum and the remaining landscape, which sets both the gap that governs the number of measurements required to identify the optimum (see Appendix~\ref{a2}) and the ranks of the resulting MPS approximation. Considering the simplistic example of a sine function, we recover an MPS approximation with maximal MPS ranks, $r = 2$. Casting this MPS into a diagonal $\mathrm{H}_{\mathrm{mpo}}$, as in Eq.~\eqref{eq7}, we notice that the MPO cores naturally admit the unitary criterion with the addition of a single ancillary qubit. In Fig.~\ref{Fig3}, we illustrate our results for $\sin\left(x\right)$ with powers $p \in \left\{ 1,10,50,100 \right\}$.

Note that, when considering the sine function, we avoid performing a search for the unitary MPO: the reshaped cores of the diagonal MPO are exactly proportional to isometries, so that an exact unitary MPO is obtained analytically by completing each rescaled core to a unitary (we verified numerically that the resulting overall scale factor is optimal, $c^{\ast} = 1$). However, for more complex functions, such as Ackley and Rosenbrock, this is not the case. Analytically recovering low-rank MPS approximations becomes challenging for these functions. For the small two-dimensional instances studied here, we obtained the MPS by truncated SVD of the (transformed) grid values; for larger problems, cross-approximation techniques recover such representations from polynomially many samples~\cite{novikov2020tensor, oseledets2010tt}. For Ackley, an MPS with maximal ranks $r = 6$ approximates the transformed objective with a relative accuracy of $5.1\times10^{-3}$, whereas for Rosenbrock, the representation with maximal ranks $r = 16$ is numerically exact (the transformed Rosenbrock function is remarkably flat, and already $r = 6$ yields a relative accuracy of $1.7\times10^{-4}$).

\begin{figure}[ht!]
    \centering
    \includegraphics[width = \columnwidth]{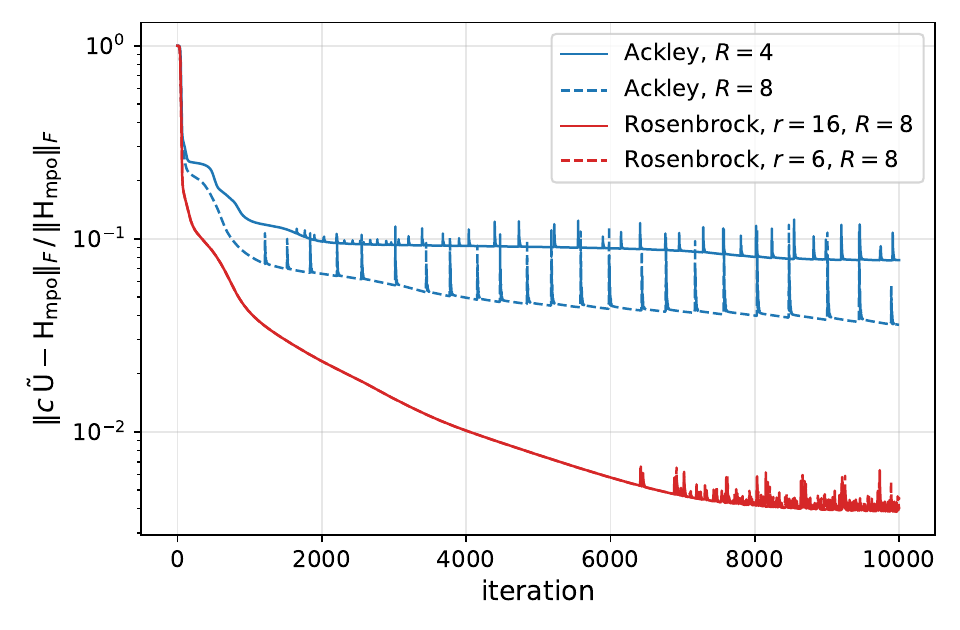}
    \caption{Convergence of the Riemannian minimization of Eq.~\eqref{eq9}: relative Frobenius error of the unitary MPO approximation versus the number of gradient-descent iterations (learning rate $0.02$, random Stiefel-manifold initialization).}
    \label{Fig:conv}
\end{figure}

\begin{figure*}[ht!]
    \centering
    \includegraphics[width = \textwidth]{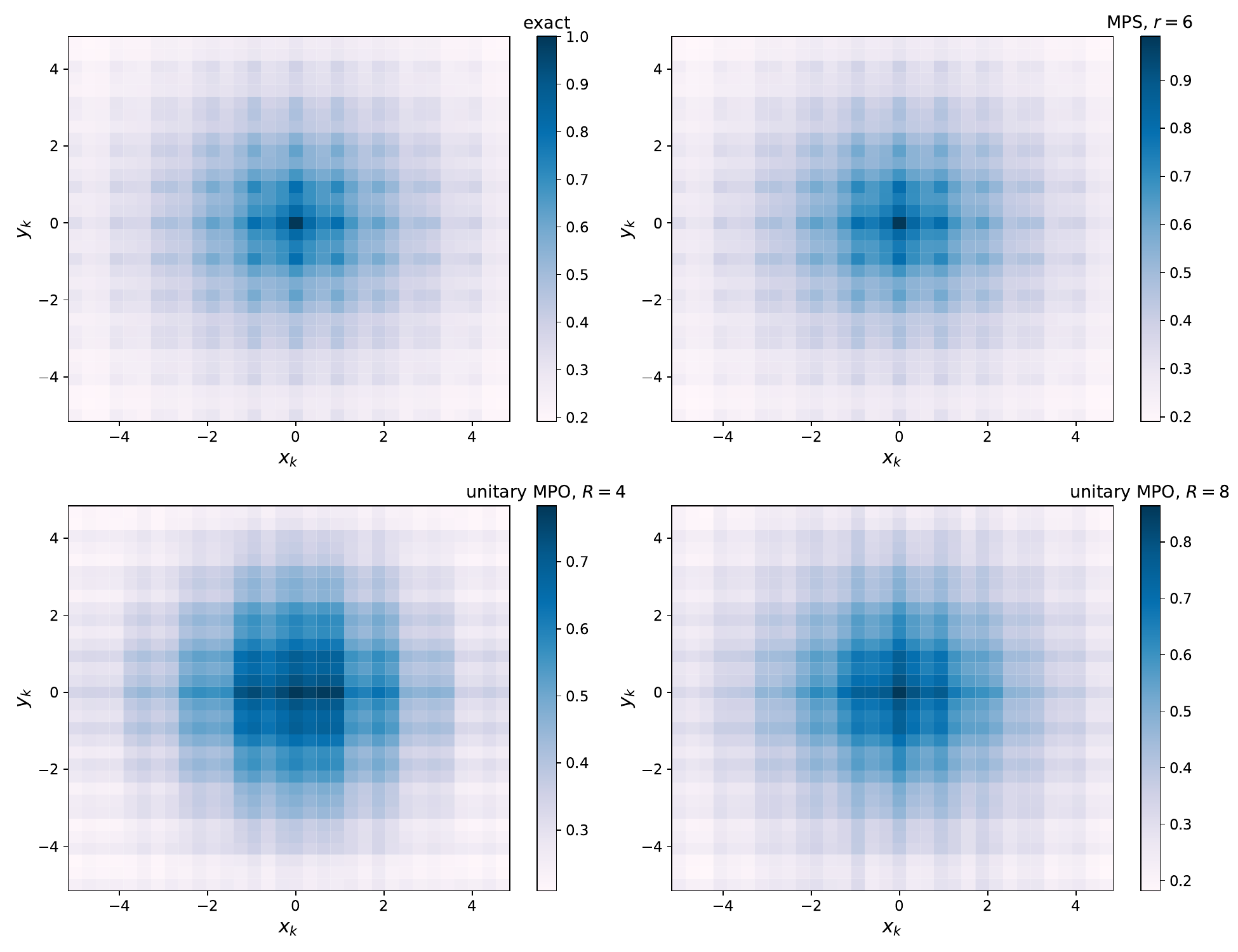}
    \caption{Comparing exact $f\left(x_k,y_k\right)$ with low-rank tensor network approximations for the two-dimensional Ackley function with 10 qubits in total (5 per dimension) for domain discretization.
    (Top left) represents the exact two-dimensional Ackley function with where the global optimum is attained at $x,y = 0$. 
    (Top right) represents the MPS approximation with maximal ranks, $r = 6$. 
    The bottom row represents the reconstructed function from the unitary MPO obtained by minimizing Eq.~\eqref{eq9} for the following cases: (Bottom left) MPO ranks $R = 4$ and (Bottom right) $R = 8$.}
    \label{Fig4}
\end{figure*}

In general, to obtain good approximations to the target MPO by minimizing Eq.~\eqref{eq9}, one has to consider larger ranks, $R \geq r$, for the unitary MPO ansatz. For the Ackley function and for Rosenbrock, we see that with increasing ranks, the quality of our approximations also improves. However, rank-deficient and rank-excessive cases may also be of interest (for an illustrative example, see Fig.~\ref{Fig4}). We observe that in all cases, Eq.~\eqref{eq9} can indeed be minimized by Riemannian gradient descent with a learning rate of $0.02$ and a total of $10^4$ iterations. The convergence of the relative Frobenius error, $\norm{c \cdot \tilde{\mathrm{U}} - \mathrm{H}_{\mathrm{mpo}}}_{F} / \norm{\mathrm{H}_{\mathrm{mpo}}}_{F}$, is shown in Fig.~\ref{Fig:conv}: starting from random initializations on the Stiefel manifold, the optimization decreases the cost monotonically and reaches $7.7\%$ ($R=4$) and $3.6\%$ ($R=8$) for Ackley, and $0.39\%$ ($R=8$) for Rosenbrock, within $10^4$ iterations. Each gradient step costs $\mathcal{O}\left(nR^3\right)$ (see Table~\ref{tab:1}); as for any non-convex optimization, no global convergence guarantee can be given, and the reached accuracy depends on the initialization: across three random restarts, the $R=4$ Ackley fit attained relative errors of $7.7\%$, $10.4\%$ and $14.5\%$. We therefore adopt the standard protocol of performing several restarts and retaining the solution with the smallest cost --- a purely classical selection step. Instructively, the restarts themselves probe the reliability criterion discussed below: the two restarts with errors below the gap of the transformed objective ($\delta \approx 0.17$) correctly identify the optimum at $p = 25$, whereas the restart with error comparable to the gap ($14.5\%$) loses it. A summary of all quantitative diagnostics --- approximation errors, scale factors, gaps, post-selection probabilities and compiled gate counts --- is provided in Table~\ref{tab:2}.

We observe that the recovered $\mathrm{U}_{\mathrm{mpo}}$ approximates the target $\mathrm{H}_{\mathrm{mpo}}$ adequately. However,  the compilation step, needed to decompose $4$ -qubit and $5$ -qubit operations into a sequence of two- and single-qubit gates, results in prohibitively deep circuits. Therefore, we are forced to consider quantum power iterations with a rank-deficient unitary MPO. In this case, convergence to the global optimum cannot be guaranteed. Yet, the approach is still motivated by the fact that power iterations find the maximal values as prescribed by Eq.~\eqref{eq4}. Even if a rank-deficient MPO admits large optimization errors, if the maxima are resolved accurately, then the unitary MPO, $\mathrm{U}_{\mathrm{mpo}}$, may still be viable. 

\begin{figure*}[ht!]
    \centering
    \includegraphics[width = 0.98\textwidth]{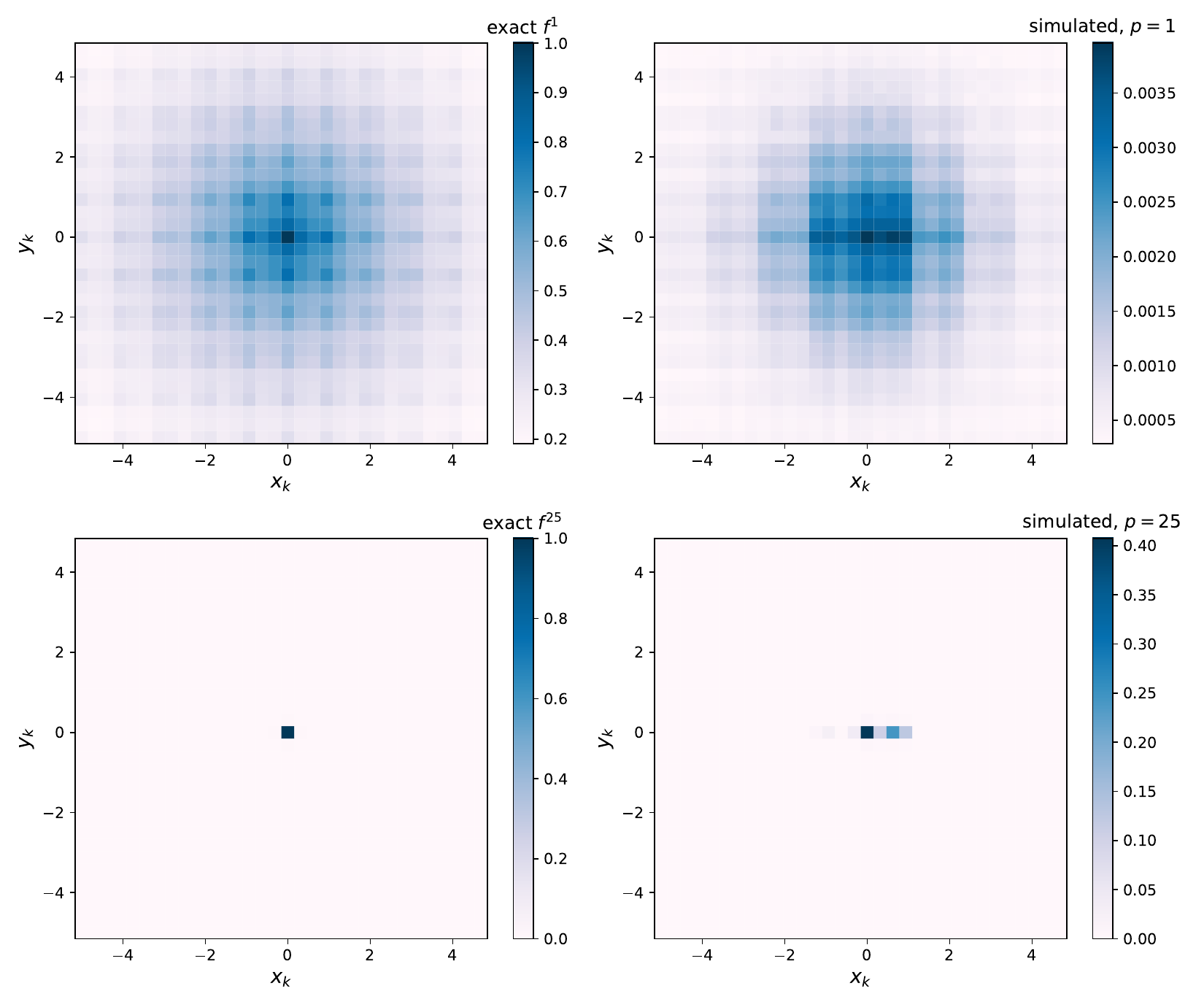}
    \caption{Quantum power iterations for the two-dimensional Ackley function. 
    The left panels indicate the powers of the exact function, $f^{j}\left(x_k,y_k\right), ~ j\in \{1,25\}$, and the right panels indicate the measurement probabilities obtained by simulating quantum power iterations. Note that when constructing the circuit, a rank-deficient $R=4$ unitary MPO is employed. Although minimization of Eq.~\eqref{eq9} achieves a worse approximation compared to the rank-excessive $R = 8$ unitary MPO (see Fig.~\ref{Fig4}), the global optimum is still resolved. 
    Therefore, quantum power iterations still converges to the global optimum of the exact function.}
    \label{Fig5} 
\end{figure*}

In Fig.~\ref{Fig4}, we compare exact versus low-rank approximations for the Ackley function and observe the following: (1) both the MPS and the $R=8$ unitary MPO qualitatively recover the exact function landscape; and (2) the rank-deficient $R=4$ unitary MPO indeed resolves the optima of the exact function. In Fig.~\ref{Fig5}, we illustrate our results on performing quantum power iterations with a rank-deficient $R=4$ unitary MPO. As expected, we observe that quantum power iterations converge to the global optimum of the Ackley function: the success of this benchmark is explained by its large relative gap, $\delta = (\tilde{f}_{(1)} - \tilde{f}_{(2)})/\tilde{f}_{(1)} \approx 0.17$, where $\tilde{f}_{(1)}$ and $\tilde{f}_{(2)}$ denote the largest and second-largest discretized values of the transformed objective. This gap exceeds the unitary MPO approximation errors ($7.7\%$ for $R = 4$) by more than a factor of two, so that the ordering of the top function values is preserved by the approximate circuit.

\begin{figure*}[ht!]
    \centering
    \includegraphics[width = \textwidth]{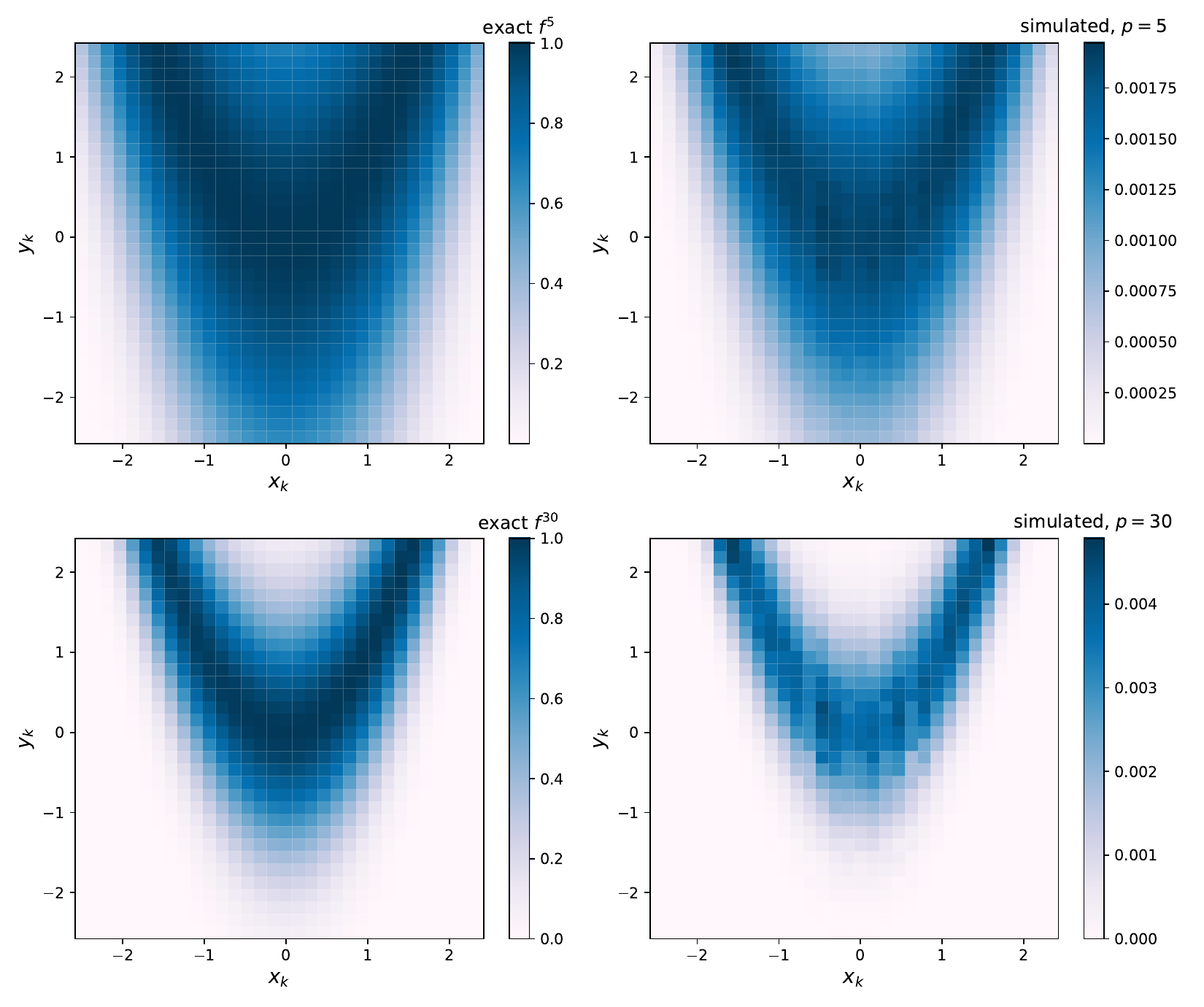}
    \caption{Quantum power iterations for the two-dimensional Rosenbrock function. 
    The left panels indicate the powers of the exact function, $f^{j}\left(x_k,y_k\right), ~ j\in \{5,30\}$, and the right panels indicate the measurement probabilities obtained via simulations. Note that when constructing the circuit, a rank-restricted $R=8$ unitary MPO is employed. With increasing powers, we observe that the banana valley is faithfully reproduced.}
    \label{Fig6} 
\end{figure*}

For the case of the Rosenbrock function, the situation is qualitatively different, and a quantitative analysis reveals the source of the difficulty. Although the unitary MPO fitting errors are small ($0.39\%$ relative Frobenius error at $R = 8$), the global optimum is not accurately resolved: quantum power iterations concentrate the measurement probability inside the banana-shaped valley, but on a grid point different from the true discretized optimum (see Fig.~\ref{Fig6}, where the valley is nevertheless faithfully reproduced with increasing powers). The reason is the flatness of the transformed objective: with the pre-processing of Eq.~\eqref{eq:preproc} and $\kappa = f\left(a,a\right) \approx 7.7\times10^{3}$, the relative gap of the discretized Rosenbrock function is $\delta \approx 8.9\times10^{-7}$ --- more than three orders of magnitude \emph{below} the achieved approximation error. In this regime, the ordering of the top values cannot be preserved by the approximate circuit, and no realistic number of power iterations can separate them ($p \sim 1/\delta$ iterations would be required). Notably, this failure is \emph{not} primarily a rank-deficiency effect: the MPS representation is numerically exact at $r = 16$, and even the modest $r = 6$ approximation carries an error ($1.7\times10^{-4}$) far smaller than the fitting error. We also note that, at 5 qubits per dimension, the continuum optimum $\left(1,1\right)$ does not lie on the discretization grid; the discretized optimum is located at $\left(1.25, 1.5625\right)$, where $y = x^2$ holds exactly on the grid. This analysis points to the quantity that actually governs reliable operation: the operator encoded by the circuit must preserve the \emph{location} of the largest discretized value, which requires the relative gap $\delta$ of the (pre-processed and discretized) objective to be large compared with the errors incurred in the encoding. A rigorous, if conservative, sufficient condition is available: if the encoded diagonal deviates elementwise from the discretized objective by less than half the absolute gap, $\max_k \lvert c^{\ast} \tilde{\mathrm{U}}_{kk} - \tilde{f}\left(x_k\right) \rvert < ( \tilde{f}_{(1)} - \tilde{f}_{(2)} )/2$, then the largest encoded diagonal value is attained at the true discretized optimum. Since the Frobenius norm upper-bounds any single-entry deviation, the reported error provides a sufficient certificate whenever $\varepsilon_{\mathrm{fit}} \cdot \norm{\mathrm{H}_{\mathrm{mpo}}}_F < ( \tilde{f}_{(1)} - \tilde{f}_{(2)} )/2$. In practice this certificate is very conservative, as the fitting error spreads over all $4^n$ matrix elements rather than concentrating on the top diagonal entries.

It is important to recognize what this condition does \emph{not} establish. It constrains only the \emph{diagonal} of the encoded operator, whereas the circuit applies the full block $\hat{\mathrm{H}} = c^{\ast}\tilde{\mathrm{U}}$ repeatedly, and $\hat{\mathrm{H}}$ also carries off-diagonal errors; preservation of the largest diagonal entry does not by itself control the $p$-fold iteration. To close this gap, we decompose $\hat{\mathrm{H}} = \mathrm{D} + \mathrm{N}$ into its diagonal and off-diagonal parts. If $\mathrm{N} = 0$ the iteration is exactly diagonal, the argmax is preserved at every power, and the half-gap condition is all that is required. For $\mathrm{N} \neq 0$ the appropriate statement is $p$-independent, since power iteration converges to the dominant eigenvector $\ket{v_1}$ of $\hat{\mathrm{H}}$. Regarding $\mathrm{N}$ as a perturbation of $\mathrm{D}$, whose dominant eigenvector is exactly $\ket{k^{\ast}}$ whenever the half-gap condition above holds, Weyl's inequality places the dominant eigenvalue of $\hat{\mathrm{H}}$ within $\norm{\mathrm{N}}_{2}$ of $d_{(1)}$; if the two largest diagonal entries $d_{(1)} > d_{(2)}$ of $\mathrm{D}$ satisfy $d_{(1)} - d_{(2)} > \norm{\mathrm{N}}_{2}$, the Davis--Kahan theorem then bounds the angle $\theta$ between $\ket{v_1}$ and $\ket{k^{\ast}}$ by
\begin{equation}\label{eq:dk}
    \sin\theta \;\leq\; \dfrac{\norm{\mathrm{N}}_{2}}{d_{(1)} - d_{(2)} - \norm{\mathrm{N}}_{2}}.
\end{equation}
Since $\sin\theta < 1/\sqrt{2}$ forces $\lvert \braket{k^{\ast}}{v_1} \rvert^2 > 1/2$ and hence exceeds every other basis-state weight, this certifies that the measured argmax is $k^{\ast}$ at large $p$. Equation~\eqref{eq:dk} is stated for Hermitian $\hat{\mathrm{H}}$; the fitted blocks are Hermitian only up to the small residual reported below, so we regard it as a guide whose conclusion we verify directly by computing $\ket{v_1}$. Unlike the diagonal condition, Eq.~\eqref{eq:dk} does control the repeated application of the full approximate operator.

We evaluated these quantities for every fitted circuit; they are collected in Table~\ref{tab:3}. For the benchmark fits of Table~\ref{tab:2} the off-diagonal part is small in Frobenius weight, between $0.1\%$ and $1.5\%$ of $\norm{\tilde{\mathrm{U}}}_F$, and the blocks are nearly Hermitian ($\norm{\hat{\mathrm{H}} - \hat{\mathrm{H}}^{\dagger}}_F / \norm{\hat{\mathrm{H}}}_F$ between $0.2\%$ and $2.3\%$). The $\exp\left(-f\right)$ fit discussed below is the exception, with $24\%$ off-diagonal weight and $40\%$ non-Hermiticity --- and it nevertheless recovers the optimum, which is itself informative: what decides the outcome is the encoded diagonal, not the size of the off-diagonal part. For the $R = 8$ Ackley fit the certificate is met outright: $d_{(1)} - d_{(2)} = 6.7\times10^{-2}$ against $\norm{\mathrm{N}}_{2} = 1.7\times10^{-2}$, so that Eq.~\eqref{eq:dk} gives $\sin\theta \leq 0.35$; the numerically computed overlap of the dominant eigenvector with $\ket{k^{\ast}}$ is $0.985$, comfortably above the certified $1 - \sin^2\theta = 0.88$. In the other successful runs the diagonal separation is smaller than $\norm{\mathrm{N}}_{2}$, so the bound --- being sufficient but not necessary --- is inconclusive even though the method succeeds; the directly computed overlaps are nevertheless at least $0.97$ throughout (Table~\ref{tab:3}). As a direct control we repeated every power iteration with the off-diagonal part deleted: for all Ackley instances the diagonal-only and full-block iterations return the same grid point, whereas for Rosenbrock they differ (indices $223$ and $863$ respectively, against the true discretized optimum $794$) --- the single case in which the objective gap lies below the off-diagonal error scale, exactly as the criterion anticipates.

These diagnostics also lead us to withdraw the empirical threshold stated in the previous version of this work, which related success to the ratio $\varepsilon_{\mathrm{fit}}/\delta$ of the relative Frobenius error to the relative gap. Further experiments produced counterexamples on both sides. With the sharper $\tilde{f} = \exp\left(-f\right)$ pre-processing of the Ackley function the fit is very poor relative to the gap ($\varepsilon_{\mathrm{fit}} = 0.90$, $\delta = 0.89$, ratio $1.01$) yet the optimum is recovered at every power studied; conversely, a scale-constrained $R = 4$ Ackley fit (see below) has a comfortable ratio of $0.47$ and nonetheless loses the optimum. The global Frobenius error is therefore not a reliable predictor in either direction, which is unsurprising given that it aggregates $4^n$ matrix elements while the outcome is decided by a few entries near the top of the spectrum. Across all of our experiments the reliable indicator is the encoded diagonal itself: the power iterations recover $k^{\ast}$ precisely when $\arg\max_{k} \hat{\mathrm{H}}_{kk} = k^{\ast}$, with the off-diagonal part controlled as above.

 The gap itself is controlled by the pre-processing, which mediates a trade-off: we verified numerically that decreasing $\kappa$ in Eq.~\eqref{eq:preproc} (or, alternatively, employing the sharper transformation $\tilde{f} = \exp\left(-f\right)$, whose gap is $\delta \approx 6.8\times10^{-3}$) does increase the gap, but simultaneously renders the transformed objective harder to encode --- the unitary MPO fitting error grows faster than the gap at fixed ansatz rank ($39\%$ relative error for $\exp\left(-f\right)$ even at $R = 16$), so that within the studied family of pre-processings and ansatz ranks the Rosenbrock optimum remains unresolved. Overcoming this trade-off, for instance via exact block-encoding constructions that bypass the variational fit~\cite{nibbi2023block}, is an interesting direction for future work.

In all the cases that were studied, we compiled the unitary MPO into a Qiskit gate sequence of single- and two-qubit gates~\cite{Qiskit,Rakyta2022approaching}, from which the gate counts and depths reported below are obtained. We verified that the compiled sequences do reproduce the cores they were derived from: every transpiled core agrees with its target elementwise to better than $3\times10^{-13}$, the only exception being the analytically constructed sine circuit, whose two-qubit cores are synthesized by Qiskit's Cartan decomposition with a residual of $2\times10^{-5}$. The measurement probabilities displayed in Figs.~\ref{Fig3}, \ref{Fig5} and \ref{Fig6} were obtained by exact state-vector evolution under the encoded block $\tilde{\mathrm{U}}$, cascading it the appropriate number of times and post-selecting the ancillary registers on the all-zero outcome after every iteration; this is equivalent to simulating the encoded circuit and avoids any sampling error.

To probe the behavior of the method beyond two dimensions, we additionally consider the four-dimensional Ackley function on $\left[-5,5\right]^4$, discretized with 5 qubits per dimension ($n = 20$ system qubits, $2^{20} \approx 10^{6}$ grid points), with the same pre-processing as in Eq.~\eqref{eq:preproc}. An MPS with maximal ranks $r = 8$ approximates the transformed objective with a relative accuracy of $1.8\times10^{-3}$, and the relative gap is $\delta \approx 0.11$. With a rank-restricted $R = 8$ unitary MPO ($6.0\%$ fitting error for the best of three restarts) and, more robustly, with $R = 16$ ($4.0\%$), quantum power iterations correctly identify the global optimum at the origin for all studied powers $p \leq 25$. The restart variability again tracks the encoded diagonal rather than the global fit error: the two $R = 8$ restarts that fail are precisely those whose encoded diagonal attains its maximum away from $k^{\ast}$.

Table~\ref{tab:2} collects the quantitative diagnostics for all benchmarks: the relative accuracy of the MPS representation, the relative Frobenius error of the unitary MPO fit, the magnitude of the optimal scale factor $c^{\ast}$ of Eq.~\eqref{eq9}, the relative gap $\delta$ of the discretized objective, the CNOT count and circuit depth of one compiled iteration block, and the cumulative success (post-selection) probability at the largest power studied. Two observations are worth emphasizing. First, the benchmarks succeed exactly when the encoded diagonal preserves the location of the maximum, which for these instances requires a gap comfortably larger than the errors incurred in the encoding. (The sine benchmark, with its analytically exact circuit, succeeds despite a tiny gap; note that for a smooth objective the gap to the \emph{neighboring} grid points near the optimum scales as the square of the discretization step, so a finite number of measurements would return one of several nearly-degenerate neighbors --- which, for the purpose of continuous optimization, are equally good candidate optima. The gap criterion governs the exact identification of the discretized optimum, as in Appendix~\ref{a2}.) Second, the scale factor $c^{\ast}$ has a direct physical meaning: the circuit implements the block $\tilde{\mathrm{U}} \approx \mathrm{H}_{\mathrm{mpo}}/c^{\ast}$, and hence each iteration succeeds with a probability suppressed by $1/\left(c^{\ast}\right)^2$. A fitted scale $c^{\ast} > 1$ therefore introduces an exponential-in-$p$ overhead $\left(c^{\ast}\right)^{-2p}$ on top of the idealized analysis of Appendix~\ref{a1} (which assumes $c^{\ast} = 1$). For the analytically constructed sine circuit, $c^{\ast} = 1$ is optimal and the measured success probabilities reproduce Eq.~\eqref{A:eq14} to within numerical precision; for the variational fits, $c^{\ast} \approx 2.7\textrm{--}2.9$ (Ackley) leads to a rapid decay of the success probability with $p$. In the previous version of this work we suggested constraining $c^{\ast}$ during the variational search as a possible mitigation; we have now carried this out, with the results described next.

% BEGIN AUTO SCALEPARA (generated by experiments/make_tables.py)
The scale factor is not a property of the target but an outcome of the \emph{unconstrained} minimization of Eq.~\eqref{eq9}, in which nothing rewards a small $c$. Since every block of a unitary matrix has operator norm at most one, the smallest subnormalization compatible with encoding $\mathrm{H}_{\mathrm{mpo}}$ is $c = \norm{\mathrm{H}_{\mathrm{mpo}}}_{2} = \max_{k} \tilde{f}\left(x_k\right) \approx 1$. We therefore repeated the variational search with $c$ held fixed at that value throughout, instead of resetting it to its least-squares optimum at every gradient step, warm-starting the constrained search at the corresponding unconstrained solution. Table~\ref{tab:4} collects the outcome.

The gain is large and, for the Ackley benchmarks, costs nothing in accuracy. For the two-dimensional Ackley function at $R = 8$, driving $c$ from its unconstrained value $2.75$ down to unity moves the fitting error from $3.6\%$ to $2.9\%$ --- slightly \emph{better}, since the warm-started constrained search settles in a different basin --- leaves the optimum correctly identified at every power, and raises the cumulative post-selection probability at $p = 25$ from $7.4\times10^{-29}$ to $4.3\times10^{-7}$ --- an improvement of twenty-two orders of magnitude, obtained purely by reformulating the classical fitting objective. The rank-deficient $R = 4$ ansatz behaves the same way, with the fitting error essentially flat ($7.7\%$ to $8.3\%$) and the post-selection probability rising from $7.3\times10^{-32}$ to $1.2\times10^{-9}$; intermediate values of $c$ interpolate monotonically between the extremes (at $c = 1.5$, $\varepsilon_{\mathrm{fit}} = 7.9\%$ and $P_{\mathrm{success}} = 8.1\times10^{-18}$). The four-dimensional benchmark improves likewise, from $2.1\times10^{-20}$ to $5.6\times10^{-13}$ at $R = 8$ and from $6.1\times10^{-17}$ to $1.0\times10^{-11}$ at $R = 16$, in both cases still returning the global optimum. Rosenbrock is the one case in which the constraint costs accuracy: the fitting error grows from $0.39\%$ to $4.2\%$ while the post-selection probability rises from $2.0\times10^{-4}$ to $7.0\times10^{-2}$, and its optimum remains unresolved either way for the reasons given above. The mitigation is also free in circuit resources: the compiled blocks of the constrained fits differ from the unconstrained ones by less than $0.5\%$ in CNOT count (for instance $997$ against $1000$ for the two-dimensional Ackley function at $R = 8$), since the constraint changes where the optimizer lands on the Stiefel manifold but not the geometry of the ansatz.

The improvement is not an artifact of one initialization. Repeating the $R = 8$ Ackley experiment from three independent random restarts, the unconstrained fits reach $\varepsilon_{\mathrm{fit}} = 3.6\%$--$5.4\%$ with post-selection probabilities spanning $6.1\times10^{-34}$ to $7.4\times10^{-29}$ at $p = 25$, whereas the corresponding constrained fits reach $2.9\%$--$3.1\%$ with probabilities of only $3.1\times10^{-7}$ to $4.3\times10^{-7}$ --- a gain of twenty-two to twenty-seven orders of magnitude in every case, with the optimum recovered throughout. The constrained fits are also markedly more reproducible: their post-selection probabilities vary by less than a factor of two across restarts, against five orders of magnitude for the unconstrained ones, because fixing $c$ removes the very degree of freedom along which the unconstrained optimizer is free to drift.

Two qualifications matter. First, the residual decay of $P_{\mathrm{success}}$ with $p$ is intrinsic: even at $c = \norm{\mathrm{H}_{\mathrm{mpo}}}_{2}$ every iteration retains the $\langle \tilde{f}^{2} \rangle$-type suppression of Appendix~\ref{a1}, so post-selection remains the dominant practical cost of the method. What the constrained fit removes is the \emph{additional} $\left(c^{\ast}\right)^{-2p}$ factor, which is an artifact of the fitting objective rather than a feature of the algorithm. Second, the constrained problem is harder to optimize: a cold random start can stall at the trivial stationary point $\tilde{\mathrm{U}} = 0$, where the cost equals $\norm{\mathrm{H}_{\mathrm{mpo}}}_F^2$. At $R = 4$ a cold-started constrained fit converged instead to a solution with a small relative Frobenius error ($8.1\%$, i.e. $0.47\,\delta$) that nonetheless fails to recover the optimum; this is the second of the two counterexamples to the withdrawn error-ratio criterion discussed above, and its diagnostics appear in Table~\ref{tab:3}.
% END AUTO SCALEPARA

\begin{table*}[ht!]
    \centering
    \begin{tabular}{l c c c c c c c c c c}
        \textbf{Benchmark} & $n$ & $r$ & $\varepsilon_{\mathrm{MPS}}$ & $R$ & $\varepsilon_{\mathrm{fit}}$ & $c^{\ast}$ & gap $\delta$ & CNOTs (depth)/iter. & $p_{\max}$ & $P_{\mathrm{success}}\left(p_{\max}\right)$ \\
        \hline
        Sine (analytic)   & 20 & 2  & $5.0\times10^{-15}$ & 2 & $0$                 & $1.000$ & $4.5\times10^{-12}$ & 58 (136)   & 100 & $5.6\times10^{-2}$ \\
        Ackley            & 10 & 6  & $5.1\times10^{-3}$  & 4 & $7.7\times10^{-2}$  & $2.91$  & $1.7\times10^{-1}$  & 194 (387)  & 25  & $7.3\times10^{-32}$ \\
        Ackley            & 10 & 6  & $5.1\times10^{-3}$  & 8 & $3.6\times10^{-2}$  & $2.75$  & $1.7\times10^{-1}$  & 1000 (1959) & 25  & $7.4\times10^{-29}$ \\
        Rosenbrock        & 10 & 16 & $8.2\times10^{-15}$ & 8 & $3.9\times10^{-3}$  & $1.13$  & $8.9\times10^{-7}$  & 995 (1949)  & 30  & $2.0\times10^{-4}$ \\
        Ackley (4D)       & 20 & 8  & $1.8\times10^{-3}$  & 8 & $6.0\times10^{-2}$  & $1.35$  & $1.1\times10^{-1}$  & 1998 (3923) & 25  & $2.1\times10^{-20}$ \\
        Ackley (4D)       & 20 & 8  & $1.8\times10^{-3}$  & 16 & $4.0\times10^{-2}$ & $1.29$  & $1.1\times10^{-1}$  & 8873 (17350) & 25  & $6.1\times10^{-17}$ \\
    \end{tabular}
    \caption{Quantitative diagnostics of quantum power iterations for all studied benchmarks. $n$ is the number of system qubits, $r$ the maximal MPS rank, $\varepsilon_{\mathrm{MPS}}$ the relative accuracy of the MPS representation of the (pre-processed) objective, $R$ the unitary MPO rank, $\varepsilon_{\mathrm{fit}}$ the relative Frobenius error of the minimization of Eq.~\eqref{eq9}, $c^{\ast}$ the fitted scale factor, $\delta$ the relative gap between the two largest discretized objective values, CNOTs (depth)/iter.\ the CNOT count and the circuit depth (in the \{single-qubit, CNOT\} gate set; upper bound obtained by sequential composition of the compiled cores) of a single iteration block (Qiskit compilation~\cite{Qiskit}), and $P_{\mathrm{success}}\left(p_{\max}\right)$ the cumulative post-selection success probability after $p_{\max}$ iterations. The global optimum is correctly identified in all cases except Rosenbrock, where the gap lies below the approximation error (see main text).}
    \label{tab:2}
\end{table*}

\begin{table*}[ht!]
    \centering
    \begin{tabular}{l c c c c c c c c c c}
        \textbf{Benchmark} & $R$ & seed & $\norm{\mathrm{N}}_F/\norm{\tilde{\mathrm{U}}}_F$ & $\norm{\mathrm{N}}_2$ & non-Herm. & $d_{(1)} - d_{(2)}$ & Eq.~\eqref{eq:dk} & $\arg\max \mathrm{D} = k^{\ast}$ & $\lvert \braket{k^{\ast}}{v_1} \rvert^2$ & optimum \\
        \hline
% BEGIN AUTO TABLE3ROWS (generated by experiments/make_tables.py)
Ackley 2D & 4 & 0 & $1.1\%$ & $9.5\times10^{-3}$ & $1.6\%$ & $8.9\times10^{-3}$ & --- & $\checkmark$ & $0.976$ & $\checkmark$ \\
Ackley 2D & 4 & 1 & $1.2\%$ & $1.0\times10^{-2}$ & $1.7\%$ & $2.5\times10^{-4}$ & --- & $\times$ & $0.084$ & $\times$ \\
Ackley 2D & 4 & 2 & $1.2\%$ & $9.5\times10^{-3}$ & $1.6\%$ & $6.9\times10^{-3}$ & --- & $\checkmark$ & $0.984$ & $\checkmark$ \\
Ackley 2D & 8 & 0 & $1.5\%$ & $1.7\times10^{-2}$ & $2.3\%$ & $6.7\times10^{-2}$ & $0.35$ & $\checkmark$ & $0.985$ & $\checkmark$ \\
Rosenbrock 2D & 8 & 0 & $0.1\%$ & $3.0\times10^{-3}$ & $0.2\%$ & $1.2\times10^{-3}$ & --- & $\times$ & $0.019$ & $\times$ \\
Ackley 4D & 8 & 1 & $0.9\%$ & $1.3\times10^{-2}$ & $1.4\%$ & $8.6\times10^{-3}$ & --- & $\checkmark$ & $0.989$ & $\checkmark$ \\
Ackley 4D & 16 & 0 & $0.7\%$ & $1.5\times10^{-2}$ & $0.9\%$ & $1.3\times10^{-2}$ & --- & $\checkmark$ & $0.998$ & $\checkmark$ \\
Ackley 2D, $e^{-f}$ & 4 & 0 & $24.5\%$ & $1.3\times10^{-2}$ & $40.4\%$ & $6.4\times10^{-3}$ & --- & $\checkmark$ & $0.999$ & $\checkmark$ \\
Ackley 2D, $c=1$ (cold) & 4 & 0 & $1.3\%$ & $1.2\times10^{-2}$ & $1.7\%$ & $1.8\times10^{-3}$ & --- & $\times$ & $0.002$ & $\times$ \\
% END AUTO TABLE3ROWS
    \end{tabular}
    \caption{Diagnostics controlling the repeated application of the encoded operator $\hat{\mathrm{H}} = c^{\ast}\tilde{\mathrm{U}} = \mathrm{D} + \mathrm{N}$, where $\mathrm{D}$ is diagonal. Columns: the Frobenius weight and the spectral norm of the off-diagonal part $\mathrm{N}$; the relative non-Hermiticity $\norm{\hat{\mathrm{H}} - \hat{\mathrm{H}}^{\dagger}}_F/\norm{\hat{\mathrm{H}}}_F$; the separation of the two largest entries of the encoded diagonal; the Davis--Kahan bound of Eq.~\eqref{eq:dk} on $\sin\theta$ between the dominant eigenvector $\ket{v_1}$ of $\hat{\mathrm{H}}$ and $\ket{k^{\ast}}$, which is conclusive only when $d_{(1)} - d_{(2)} > \norm{\mathrm{N}}_2$ (otherwise ``---''); whether the encoded diagonal attains its maximum at the true discretized optimum; the overlap $\lvert \braket{k^{\ast}}{v_1} \rvert^2$, obtained by applying $\hat{\mathrm{H}}$ repeatedly with renormalization until the overlap changes by less than $10^{-8}$ or $250$ steps are reached; and whether the power iterations return $k^{\ast}$. The last two rows are the counterexamples, discussed in the text, to a criterion based on the global Frobenius error: the $\exp(-f)$ pre-processing, whose fit is poor relative to the gap yet succeeds, and a cold-started scale-constrained fit, whose fit is good relative to the gap yet fails (Table~\ref{tab:4} reports the warm-started constrained fits used elsewhere). Across every run the outcome is decided by the encoded diagonal.}
    \label{tab:3}
\end{table*}

\begin{table*}[ht!]
    \centering
    \begin{tabular}{l l c c c c c}
        \textbf{Benchmark} & fit & $\lvert c \rvert$ & $\varepsilon_{\mathrm{fit}}$ & $p_{\max}$ & $P_{\mathrm{success}}$ & opt. \\
        \hline
% BEGIN AUTO TABLE4ROWS (generated by experiments/make_tables.py)
Ackley 2D, $R=4$ & unconstrained & $2.91$ & $7.7\%$ & 25 & $7.3\times10^{-32}$ & $\checkmark$ \\
 & $c = 1$ & $1.00$ & $8.3\%$ & 25 & $1.2\times10^{-9}$ & $\checkmark$ \\
\hline
Ackley 2D, $R=8$ & unconstrained & $2.75$ & $3.6\%$ & 25 & $7.4\times10^{-29}$ & $\checkmark$ \\
 & $c = 1$ & $1.00$ & $2.9\%$ & 25 & $4.3\times10^{-7}$ & $\checkmark$ \\
\hline
Ackley 4D, $R=8$ & unconstrained & $1.35$ & $6.0\%$ & 25 & $2.1\times10^{-20}$ & $\checkmark$ \\
 & $c = 1$ & $1.00$ & $5.2\%$ & 25 & $5.6\times10^{-13}$ & $\checkmark$ \\
\hline
Ackley 4D, $R=16$ & unconstrained & $1.29$ & $4.0\%$ & 25 & $6.1\times10^{-17}$ & $\checkmark$ \\
 & $c = 1$ & $1.00$ & $3.9\%$ & 25 & $1.0\times10^{-11}$ & $\checkmark$ \\
\hline
Rosenbrock 2D, $R=8$ & unconstrained & $1.13$ & $0.4\%$ & 30 & $2.0\times10^{-4}$ & $\times$ \\
 & $c = 1$ & $1.00$ & $4.2\%$ & 30 & $7.0\times10^{-2}$ & $\times$ \\
% END AUTO TABLE4ROWS
    \end{tabular}
    \caption{Removing the post-selection penalty of the scale factor. For each benchmark the unconstrained fit of Eq.~\eqref{eq9}, in which $c$ is reset to its least-squares optimum at every gradient step, is compared with a fit in which $c$ is held fixed at $\norm{\mathrm{H}_{\mathrm{mpo}}}_2 \approx 1$, the smallest subnormalization compatible with encoding $\mathrm{H}_{\mathrm{mpo}}$ in a block of a unitary. The constrained fits are warm-started at the corresponding unconstrained solution. $P_{\mathrm{success}}$ is the cumulative post-selection probability at $p_{\max}$, and the last column indicates whether the power iterations return the true discretized optimum.}
    \label{tab:4}
\end{table*}

\section{Discussion}

In this paper, we propose a quantum realization of the classical power iteration algorithm. Given a low-rank tensor network approximation of some discretized function, we design a variational minimization routine which allows one to cast such representations into quantum circuits using a few ancilla qubits and post-selection. Power iterations can then be realized by the repeated concatenation of such circuit blocks. 

A limitation of our approach comes from the fact that after each iteration, we require post-selection on the ancillary space in order to continue subsequent iterations. An alternative approach would involve initializing a new set of ancillary qubits at each subsequent iteration. In this case, only a pruning of the measurement statistics needs to be considered. However, as the size of the ancillary space increases linearly with the number of iterations, a limitation is placed on the number of qubits that can be employed for computation. To recover a successful computation, one is forced to repeat circuit evaluations multiple times on a real device. At first glance, one may question the scalability of our approach due to its probabilistic nature. However, this is not the case. We analytically show that the success probability for fixed number of iterations is independent of the system size (see Appendix~\ref{a1}). Therefore, one can freely choose the number of qubits that need to be employed for domain discretization.       
  
For clarity, we explicitly summarize the assumptions under which the proposed method can be expected to operate reliably:
(a) the objective function, after discretization and pre-processing, must admit an accurate low-rank MPS/MPO approximation~\cite{oseledets2013constructive};
(b) the variational minimization of Eq.~\eqref{eq9} must converge to a small relative Frobenius error, which requires a sufficient unitary MPO rank $R$ and a successful Riemannian optimization;
(c) the post-selection probability must remain acceptable --- while we show in Appendix~\ref{a1} that it is independent of the number of qubits, it does decay with the number of iterations $p$ at a rate set by the function landscape; and
(d) the discretized objective must exhibit a sufficient separation between the largest and the second-largest values, since the number of measurements required to identify the optimum scales inversely with the squared gap (see Appendix~\ref{a2}).
The choice of pre-processing directly influences conditions (a) and (d), as it shapes both the ranks and the gap of the transformed objective. Our Rosenbrock benchmark illustrates the interplay of conditions (b) and (d): although the unitary MPO is accurate to $0.39\%$, the pre-processed objective is so flat that its relative gap ($\approx 8.9\times10^{-7}$) lies far below this error, which prevents an accurate resolution of the global optimum.

Another practical limitation arises from the fact that present day quantum processors can only realize single- and two-qubit operations with limited connectivity. Therefore, a compilation step is necessary when casting unitary MPO cores into quantum circuits. Specifically, each constituent $\mathrm{U}_k$ of $\mathrm{U}_{\mathrm{mpo}}$, needs to be compiled into hardware realizable gate sequences. The theoretical lower-bounds for this compilation step prescribe CNOT gate counts to scale as $\mathcal{O}\left(4^m\right)$, where $m$ is the number of qubits that the gate non-trivially acts on~\cite{TheorLimQGdec}. In our method, since $U_k$ acts on $(\log_2(R)+1)$ qubits, the compilation forces the circuit depth to grow as $ \mathcal{O}\left(nR^2\right)$ with each iteration. Although numerical strategies such as those in Refs.~\cite{Rakyta2022approaching, rakyta2022efficient} indeed approximate multi-qubit gates with CNOT counts close to this theoretical lower limit, the requirements on depth may still be out of reach for modern processors. 

It is also instructive to compare our proposal against fully classical tensor network optimization. First, state-of-the-art gradient-free TT optimizers readily solve the small proof-of-principle instances studied here: on the same discretized and pre-processed benchmarks, TTOpt~\cite{sozykin2022ttopt} recovers the exact discretized optimum of both two-dimensional functions within $10^3$ function evaluations. Second, we measured the rank growth of fully classical TT power iterations (computing $\tilde{f}^{\, p}$ via repeated Hadamard products followed by TT rounding at fixed accuracy $10^{-8}$): for our benchmarks, the rounded ranks peak at moderate values (29 for Ackley, 16 for Rosenbrock) and subsequently \emph{decrease} with growing $p$, since the iterated state approaches a rank-one computational basis state. The exponential rank growth that motivates the quantum realization is therefore a worst-case phenomenon, and does not manifest for smooth, low-dimensional objectives. The potential advantage of the quantum realization lies elsewhere: the circuit depth $\mathcal{O}\left(npR^2\right)$ is oblivious to the ranks of the \emph{intermediate} iterates $\tilde{f}^{\, p}$, whereas the classical cost is governed by them. A genuine advantage thus requires function classes whose intermediate powers do not admit accurate low-rank approximations while the function itself does; identifying natural problem families in this regime --- for example, high-dimensional objectives with many well-separated near-degenerate optima --- is an important direction for future work, and we accordingly refrain from claiming a general run-time advantage for the instances studied here.

In addition to limitations, minimizing Eq.~\eqref{eq9} is another point of discussion.
Our approach relies on Riemannian gradient descent to minimize Eq.~\eqref{eq9}, and we do not have a complexity theory for this step. We can offer no bound on the number of gradient steps required, nor on the ansatz rank $R$ needed to reach a prescribed accuracy; the minimization is non-convex, so no global convergence guarantee is available, and we have seen in the seed study above that the attained accuracy depends on the initialization. This is a genuine limitation of the present work. Two considerations partially offset it. First, although we cannot certify a fit in advance, we can certify one after the fact: the encoded-diagonal condition together with Eq.~\eqref{eq:dk} decides, for a \emph{given} fitted circuit and at a cost negligible compared with the fit itself, whether the power iterations will return the discretized optimum. In the absence of an a-priori theory this is the practically relevant guarantee. Second, the variational step can be dispensed with altogether: one can avoid the optimization routine entirely and recover a block encoding analytically~\cite{nibbi2023block}, by encoding the target matrix into a block of a higher-dimensional unitary. In the same spirit, since the target operator $\mathrm{H}_{\mathrm{mpo}}$ is diagonal and sparse, block-encoding constructions tailored specifically to diagonal function oracles or sparse operators constitute a natural alternative to the general variational dilation pursued here. Prompted by this question, we have carried out the comparison quantitatively for the most natural such construction. Let $\ket{\tilde{f}} = \sum_k \tilde{f}_k \ket{k} / \norm{\tilde{f}}_2$ be the normalized state encoding the discretized objective, and let $\mathrm{V}$ be the sequential circuit that prepares it exactly from $\ket{0}^{\otimes n}$ --- a staircase of $\left(\log_2\left(r\right)+1\right)$-qubit gates obtained directly from the canonical form of the MPS~\cite{schon2005sequential, ran2020encoding, melnikov2023quantum}. On $n$ system qubits together with an $n$-qubit copy register,
\begin{equation}\label{eq:be}
    \mathrm{W} = \mathrm{CNOT}^{\otimes n}_{\mathrm{sys}\rightarrow\mathrm{copy}} \left( \mathds{1}^{\otimes n} \otimes \mathrm{V} \right)
\end{equation}
satisfies $\left(\mathds{1}^{\otimes n} \otimes \bra{0}^{\otimes n}\right) \mathrm{W} \left(\mathds{1}^{\otimes n} \otimes \ket{0}^{\otimes n}\right) = \mathrm{diag}\left(\tilde{f}\right)/\norm{\tilde{f}}_2$ exactly, because the transversal CNOTs map $\ket{k}\ket{j} \mapsto \ket{k}\ket{j \oplus k}$ so that only $j = k$ survives the post-selection. This construction is exact, requires no variational search whatsoever, and --- as we verified numerically for all our benchmarks, reproducing $\mathrm{H}_{\mathrm{mpo}}$ elementwise to better than $10^{-13}$ --- is entirely free of the unitary MPO fitting error $\varepsilon_{\mathrm{fit}}$ that limits the variational route. Both routes of course inherit the underlying MPS approximation error $\varepsilon_{\mathrm{MPS}}$, so the comparison below concerns only the additional cost of the encoding step. Its compiled cost is also competitive: $56$ CNOTs per iteration for the sine benchmark and $1991$ for the four-dimensional Ackley function, against $58$ and $1998$ for the corresponding unitary MPO circuits.

The decisive difference lies elsewhere, in the subnormalization. The construction of Eq.~\eqref{eq:be} block-encodes the $\ell_2$-normalized diagonal, so its subnormalization is $\norm{\tilde{f}}_2/\norm{\tilde{f}}_{\infty} = \norm{\mathrm{H}_{\mathrm{mpo}}}_F/\norm{\mathrm{H}_{\mathrm{mpo}}}_2$, which grows as $2^{n/2}$: it equals $12.6$ and $28.4$ for the two-dimensional benchmarks at $n = 10$, and $386$ and $724$ for the four-dimensional Ackley and the sine benchmarks at $n = 20$. The variational unitary MPO instead attains $c^{\ast} = 1.0$ to $2.9$ (Table~\ref{tab:2}), independently of $n$. Since each iteration succeeds with a probability suppressed by the square of the subnormalization, the exact route is less likely to post-select successfully by a factor $\left(\norm{\mathrm{H}_{\mathrm{mpo}}}_F/(c^{\ast}\norm{\mathrm{H}_{\mathrm{mpo}}}_2)\right)^{2}$ per iteration --- between $19$ and $5\times10^{5}$ for our benchmarks, and compounding as the $p$-th power. Equivalently, restoring an $\mathcal{O}(1)$ subnormalization by amplitude amplification costs $\mathcal{O}\left(2^{n/2}\right)$ rounds per iteration. The exact construction also requires $n + \log_2\left(r\right)$ ancillary qubits rather than $\log_2\left(R\right)$, i.e. $23$ instead of $3$ at $n = 20$. The trade-off is therefore sharp and, we believe, of independent interest: the variational dilation buys an $n$-independent subnormalization and an exponentially smaller ancilla register at the price of an approximation error, and it is this subnormalization, rather than gate count, that dominates the end-to-end cost of the iterated algorithm. A third route, block-encoding $\mathrm{diag}(\tilde{f})/\norm{\tilde{f}}_{\infty}$ directly through a coherent arithmetic oracle for $\tilde{f}$ followed by a controlled rotation, attains the optimal subnormalization but presupposes an explicit reversible circuit for the objective, which is unavailable in the black-box setting considered here. A systematic study of these alternatives, and in particular of the MPO block encodings of Ref.~\cite{nibbi2023block}, remains an interesting subject for future work. We remark that the variational route has one structural advantage: it applies verbatim to \emph{any} low-rank MPO, diagonal or not, and produces static circuit blocks of fixed, hardware-friendly geometry. For the special case of the sine function we showed that the variational search can be bypassed altogether, as the diagonal MPO cores are proportional to isometries and admit an exact analytic completion to unitaries.

\clearpage
\bibliography{refs.bib}

@article{cerezo2021variational,
  title={Variational quantum algorithms},
  author={Cerezo, Marco and Arrasmith, Andrew and Babbush, Ryan and Benjamin, Simon C and Endo, Suguru and Fujii, Keisuke and McClean, Jarrod R and Mitarai, Kosuke and Yuan, Xiao and Cincio, Lukasz and others},
  journal={Nature Reviews Physics},
  volume={3},
  number={9},
  pages={625--644},
  year={2021},
  publisher={Nature Publishing Group UK London}
}

@article{mcclean2016theory,
  title={The theory of variational hybrid quantum-classical algorithms},
  author={McClean, Jarrod R and Romero, Jonathan and Babbush, Ryan and Aspuru-Guzik, Al{\'a}n},
  journal={New Journal of Physics},
  volume={18},
  number={2},
  pages={023023},
  year={2016},
  publisher={IOP Publishing}
}

@article{bharti2022noisy,
  title={Noisy intermediate-scale quantum algorithms},
  author={Bharti, Kishor and Cervera-Lierta, Alba and Kyaw, Thi Ha and Haug, Tobias and Alperin-Lea, Sumner and Anand, Abhinav and Degroote, Matthias and Heimonen, Hermanni and Kottmann, Jakob S and Menke, Tim and others},
  journal={Reviews of Modern Physics},
  volume={94},
  number={1},
  pages={015004},
  year={2022},
  publisher={APS}
}

@article{biamonte2017quantum,
  title={Quantum machine learning},
  author={Biamonte, Jacob and Wittek, Peter and Pancotti, Nicola and Rebentrost, Patrick and Wiebe, Nathan and Lloyd, Seth},
  journal={Nature},
  volume={549},
  number={7671},
  pages={195--202},
  year={2017},
  publisher={Nature Publishing Group UK London}
}

@article{lloyd2018quantum,
  title={Quantum approximate optimization is computationally universal},
  author={Lloyd, Seth},
  journal={arXiv preprint arXiv:1812.11075},
  year={2018}
}

@article{morales2020universality,
  title={On the universality of the quantum approximate optimization algorithm},
  author={Morales, Mauro ES and Biamonte, JD and Zimbor{\'a}s, Zolt{\'a}n},
  journal={Quantum Information Processing},
  volume={19},
  number={9},
  pages={1--26},
  year={2020},
  publisher={Springer}
}

@article{zhao2023universal,
  title={A universal variational quantum eigensolver for non-Hermitian systems},
  author={Zhao, Huanfeng and Zhang, Peng and Wei, Tzu-Chieh},
  journal={Scientific Reports},
  volume={13},
  number={1},
  pages={22313},
  year={2023},
  publisher={Nature Publishing Group UK London}
}

@article{shaydulin2019hybrid,
  title={A hybrid approach for solving optimization problems on small quantum computers},
  author={Shaydulin, Ruslan and Ushijima-Mwesigwa, Hayato and Negre, Christian FA and Safro, Ilya and Mniszewski, Susan M and Alexeev, Yuri},
  journal={Computer},
  volume={52},
  number={6},
  pages={18--26},
  year={2019},
  publisher={IEEE}
}

@article{he2017decompositions,
  title={Decompositions of n-qubit Toffoli gates with linear circuit complexity},
  author={He, Yong and Luo, Ming-Xing and Zhang, E and Wang, Hong-Ke and Wang, Xiao-Feng},
  journal={International Journal of Theoretical Physics},
  volume={56},
  pages={2350--2361},
  year={2017},
  publisher={Springer}
}

@article{low2019hamiltonian,
  title={Hamiltonian simulation by qubitization},
  author={Low, Guang Hao and Chuang, Isaac L},
  journal={Quantum},
  volume={3},
  pages={163},
  year={2019},
  publisher={Verein zur F{\"o}rderung des Open Access Publizierens in den Quantenwissenschaften}
}

@article{childs2012hamiltonian,
  title={Hamiltonian simulation using linear combinations of unitary operations},
  author={Childs, Andrew M and Wiebe, Nathan},
  journal={arXiv preprint arXiv:1202.5822},
  year={2012}
}

@article{orus2014practical,
  title={A practical introduction to tensor networks: Matrix product states and projected entangled pair states},
  author={Or{\'u}s, Rom{\'a}n},
  journal={Annals of physics},
  volume={349},
  pages={117--158},
  year={2014},
  publisher={Elsevier}
}

@article{cirac2021matrix,
  title={Matrix product states and projected entangled pair states: Concepts, symmetries, theorems},
  author={Cirac, J Ignacio and Perez-Garcia, David and Schuch, Norbert and Verstraete, Frank},
  journal={Reviews of Modern Physics},
  volume={93},
  number={4},
  pages={045003},
  year={2021},
  publisher={APS}
}

@article{eisert2008area,
  title={Area laws for the entanglement entropy-a review},
  author={Eisert, Jens and Cramer, Marcus and Plenio, Martin B},
  journal={arXiv preprint arXiv:0808.3773},
  year={2008}
}

@article{gray2021hyper,
  title={Hyper-optimized tensor network contraction},
  author={Gray, Johnnie and Kourtis, Stefanos},
  journal={Quantum},
  volume={5},
  pages={410},
  year={2021},
  publisher={Verein zur F{\"o}rderung des Open Access Publizierens in den Quantenwissenschaften}
}

@inproceedings{liu2021closing,
  title={Closing the" quantum supremacy" gap: achieving real-time simulation of a random quantum circuit using a new sunway supercomputer},
  author={Liu, Yong and Liu, Xin and Li, Fang and Fu, Haohuan and Yang, Yuling and Song, Jiawei and Zhao, Pengpeng and Wang, Zhen and Peng, Dajia and Chen, Huarong and others},
  booktitle={Proceedings of the International Conference for High Performance Computing, Networking, Storage and Analysis},
  pages={1--12},
  year={2021}
}

@article{soley2021iterative,
  title={Iterative power algorithm for global optimization with quantics tensor trains},
  author={Soley, Micheline B and Bergold, Paul and Batista, Victor S},
  journal={Journal of Chemical Theory and Computation},
  volume={17},
  number={6},
  pages={3280--3291},
  year={2021},
  publisher={ACS Publications}
}

@article{dolgov2020approximation,
  title={Approximation and sampling of multivariate probability distributions in the tensor train decomposition},
  author={Dolgov, Sergey and Anaya-Izquierdo, Karim and Fox, Colin and Scheichl, Robert},
  journal={Statistics and Computing},
  volume={30},
  pages={603--625},
  year={2020},
  publisher={Springer}
}

@article{novikov2015tensorizing,
  title={Tensorizing neural networks},
  author={Novikov, Alexander and Podoprikhin, Dmitrii and Osokin, Anton and Vetrov, Dmitry P},
  journal={Advances in neural information processing systems},
  volume={28},
  year={2015}
}

@article{shor1999polynomial,
  title={Polynomial-time algorithms for prime factorization and discrete logarithms on a quantum computer},
  author={Shor, Peter W},
  journal={SIAM review},
  volume={41},
  number={2},
  pages={303--332},
  year={1999},
  publisher={SIAM}
}

@article{brassard2002quantum,
  title={Quantum amplitude amplification and estimation},
  author={Brassard, Gilles and Hoyer, Peter and Mosca, Michele and Tapp, Alain},
  journal={Contemporary Mathematics},
  volume={305},
  pages={53--74},
  year={2002},
  publisher={Providence, RI; American Mathematical Society; 1999}
}

@article{harrow2009quantum,
  title={Quantum algorithm for linear systems of equations},
  author={Harrow, Aram W and Hassidim, Avinatan and Lloyd, Seth},
  journal={Physical review letters},
  volume={103},
  number={15},
  pages={150502},
  year={2009},
  publisher={APS}
}

@article{plesch2011quantum,
  title={Quantum-state preparation with universal gate decompositions},
  author={Plesch, Martin and Brukner, {\v{C}}aslav},
  journal={Physical Review A},
  volume={83},
  number={3},
  pages={032302},
  year={2011},
  publisher={APS}
}

@article{khoromskij2011d,
  title={O (d log N)-quantics approximation of N-d tensors in high-dimensional numerical modeling},
  author={Khoromskij, Boris N},
  journal={Constructive Approximation},
  volume={34},
  pages={257--280},
  year={2011},
  publisher={Springer}
}

@article{perez2006matrix,
  title={Matrix product state representations},
  author={Perez-Garcia, David and Verstraete, Frank and Wolf, Michael M and Cirac, J Ignacio},
  journal={arXiv preprint quant-ph/0608197},
  year={2006}
}

@article{glasser2019expressive,
  title={Expressive power of tensor-network factorizations for probabilistic modeling},
  author={Glasser, Ivan and Sweke, Ryan and Pancotti, Nicola and Eisert, Jens and Cirac, Ignacio},
  journal={Advances in neural information processing systems},
  volume={32},
  year={2019}
}

@article{ballani2014tree,
  title={Tree adaptive approximation in the hierarchical tensor format},
  author={Ballani, Jonas and Grasedyck, Lars},
  journal={SIAM journal on scientific computing},
  volume={36},
  number={4},
  pages={A1415--A1431},
  year={2014},
  publisher={SIAM}
}

@article{pfeifer2014faster,
  title={Faster identification of optimal contraction sequences for tensor networks},
  author={Pfeifer, Robert NC and Haegeman, Jutho and Verstraete, Frank},
  journal={Physical Review E},
  volume={90},
  number={3},
  pages={033315},
  year={2014},
  publisher={APS}
}

@article{zhang2022quantum,
  title={Quantum state preparation with optimal circuit depth: Implementations and applications},
  author={Zhang, Xiao-Ming and Li, Tongyang and Yuan, Xiao},
  journal={Physical Review Letters},
  volume={129},
  number={23},
  pages={230504},
  year={2022},
  publisher={APS}
}

@article{schon2005sequential,
  title={Sequential generation of entangled multiqubit states},
  author={Sch{\"o}n, Christian and Solano, Enrique and Verstraete, Frank and Cirac, J Ignacio and Wolf, Michael M},
  journal={Physical review letters},
  volume={95},
  number={11},
  pages={110503},
  year={2005},
  publisher={APS}
}

@article{schon2007sequential,
  title={Sequential generation of matrix-product states in cavity QED},
  author={Sch{\"o}n, C and Hammerer, K and Wolf, Michael M and Cirac, J Ignacio and Solano, E},
  journal={Physical Review A},
  volume={75},
  number={3},
  pages={032311},
  year={2007},
  publisher={APS}
}

@article{schwarz2012preparing,
  title={Preparing projected entangled pair states on a quantum computer},
  author={Schwarz, Martin and Temme, Kristan and Verstraete, Frank},
  journal={Physical review letters},
  volume={108},
  number={11},
  pages={110502},
  year={2012},
  publisher={APS}
}

@article{lubasch2020variational,
  title={Variational quantum algorithms for nonlinear problems},
  author={Lubasch, Michael and Joo, Jaewoo and Moinier, Pierre and Kiffner, Martin and Jaksch, Dieter},
  journal={Physical Review A},
  volume={101},
  number={1},
  pages={010301},
  year={2020},
  publisher={APS}
}

@article{markov2008simulating,
  title={Simulating quantum computation by contracting tensor networks},
  author={Markov, Igor L and Shi, Yaoyun},
  journal={SIAM Journal on Computing},
  volume={38},
  number={3},
  pages={963--981},
  year={2008},
  publisher={SIAM}
}

@article{preskill2018quantum,
  title={Quantum computing in the {NISQ} era and beyond},
  author={Preskill, John},
  journal={Quantum},
  volume={2},
  pages={79},
  year={2018},
  publisher={Verein zur F{\"o}rderung des Open Access Publizierens in den Quantenwissenschaften}
}

@article{melnikov2023quantum,
  title={Quantum state preparation using tensor networks},
  author={Melnikov, Artem A and Termanova, Alena A and Dolgov, Sergey V and Neukart, Florian and Perelshtein, Michael},
  journal={Quantum Science and Technology},
  volume={8},
  number={3},
  pages={035027},
  year={2023},
  publisher={IOP Publishing}
}

@article{ran2020encoding,
  title={Encoding of matrix product states into quantum circuits of one-and two-qubit gates},
  author={Ran, Shi-Ju},
  journal={Physical Review A},
  volume={101},
  number={3},
  pages={032310},
  year={2020},
  publisher={APS}
}

@article{childs2021high,
  title={High-precision quantum algorithms for partial differential equations},
  author={Childs, Andrew M and Liu, Jin-Peng and Ostrander, Aaron},
  journal={Quantum},
  volume={5},
  pages={574},
  year={2021},
  publisher={Verein zur F{\"o}rderung des Open Access Publizierens in den Quantenwissenschaften}
}

@article{hou1978cubic,
  title={Cubic splines for image interpolation and digital filtering},
  author={Hou, Hsieh and Andrews, H},
  journal={IEEE Transactions on acoustics, speech, and signal processing},
  volume={26},
  number={6},
  pages={508--517},
  year={1978},
  publisher={IEEE}
}

@article{cichocki2016tensor,
  title={Tensor networks for dimensionality reduction and large-scale optimization: Part 1 low-rank tensor decompositions},
  author={Cichocki, Andrzej and Lee, Namgil and Oseledets, Ivan and Phan, Anh-Huy and Zhao, Qibin and Mandic, Danilo P and others},
  journal={Foundations and Trends{\textregistered} in Machine Learning},
  volume={9},
  number={4-5},
  pages={249--429},
  year={2016},
  publisher={Now Publishers, Inc.}
}

@article{huggins2019towards,
  title={Towards quantum machine learning with tensor networks},
  author={Huggins, William and Patil, Piyush and Mitchell, Bradley and Whaley, K Birgitta and Stoudenmire, E Miles},
  journal={Quantum Science and technology},
  volume={4},
  number={2},
  pages={024001},
  year={2019},
  publisher={IOP Publishing}
}

@inproceedings{arasu2002pagerank,
  title={PageRank computation and the structure of the web: Experiments and algorithms},
  author={Arasu, Arvind and Novak, Jasmine and Tomkins, Andrew and Tomlin, John},
  booktitle={Proceedings of the eleventh international World Wide Web conference, poster track},
  pages={107--117},
  year={2002}
}

@article{panza2018application,
  title={Application of power method and dominant eigenvector/eigenvalue concept for approximate eigenspace solutions to mechanical engineering algebraic systems},
  author={Panza, Michael J},
  journal={American Journal of Mechanical Engineering},
  volume={6},
  number={3},
  pages={98--113},
  year={2018}
}

@book{golub2013matrix,
  title={Matrix computations},
  author={Golub, Gene H and Van Loan, Charles F},
  year={2013},
  publisher={JHU press}
}

@book{jung2001methode,
  title={Methode der finiten Elemente f{\"u}r Ingenieure},
  author={Jung, Michael and Langer, Ulrich},
  year={2001},
  publisher={Springer}
}

@article{oseledets2011tensor,
  title={Tensor-train decomposition},
  author={Oseledets, Ivan V},
  journal={SIAM Journal on Scientific Computing},
  volume={33},
  number={5},
  pages={2295--2317},
  year={2011},
  publisher={SIAM}
}

@article{sozykin2022ttopt,
  title={TTOpt: A maximum volume quantized tensor train-based optimization and its application to reinforcement learning},
  author={Sozykin, Konstantin and Chertkov, Andrei and Schutski, Roman and Phan, Anh-Huy and Cichocki, Andrzej S and Oseledets, Ivan},
  journal={Advances in Neural Information Processing Systems},
  volume={35},
  pages={26052--26065},
  year={2022}
}

@article{biamonte2017tensor,
  title={Tensor networks in a nutshell},
  author={Biamonte, Jacob and Bergholm, Ville},
  journal={arXiv preprint arXiv:1708.00006},
  year={2017}
}

@article{oseledets2013constructive,
  title={Constructive representation of functions in low-rank tensor formats},
  author={Oseledets, Ivan V},
  journal={Constructive Approximation},
  volume={37},
  pages={1--18},
  year={2013},
  publisher={Springer}
}

@inproceedings{huber2017randomized,
  title={A randomized tensor train singular value decomposition},
  author={Huber, Benjamin and Schneider, Reinhold and Wolf, Sebastian},
  booktitle={Compressed Sensing and its Applications: Second International MATHEON Conference 2015},
  pages={261--290},
  year={2017},
  organization={Springer}
}

@article{oseledets2010tt,
  title={TT-cross approximation for multidimensional arrays},
  author={Oseledets, Ivan and Tyrtyshnikov, Eugene},
  journal={Linear Algebra and its Applications},
  volume={432},
  number={1},
  pages={70--88},
  year={2010},
  publisher={Elsevier}
}

@article{luchnikov2021riemannian,
  title={Riemannian geometry and automatic differentiation for optimization problems of quantum physics and quantum technologies},
  author={Luchnikov, Ilia A and Krechetov, Mikhail E and Filippov, Sergey N},
  journal={New Journal of Physics},
  volume={23},
  number={7},
  pages={073006},
  year={2021},
  publisher={IOP Publishing}
}

@article{kressner2016preconditioned,
  title={Preconditioned low-rank Riemannian optimization for linear systems with tensor product structure},
  author={Kressner, Daniel and Steinlechner, Michael and Vandereycken, Bart},
  journal={SIAM Journal on Scientific Computing},
  volume={38},
  number={4},
  pages={A2018--A2044},
  year={2016},
  publisher={SIAM}
}

@article{rakyta2022efficient,
  title={Efficient quantum gate decomposition via adaptive circuit compression},
  author={Rakyta, P{\'e}ter and Zimbor{\'a}s, Zolt{\'a}n},
  journal={arXiv preprint arXiv:2203.04426},
  year={2022}
}

@article{novikov2020tensor,
  title={Tensor train decomposition on tensorflow (t3f)},
  author={Novikov, Alexander and Izmailov, Pavel and Khrulkov, Valentin and Figurnov, Michael and Oseledets, Ivan},
  journal={The Journal of Machine Learning Research},
  volume={21},
  number={1},
  pages={1105--1111},
  year={2020},
  publisher={JMLRORG}
}

@article{nibbi2023block,
  title={Block encoding of matrix product operators},
  author={Nibbi, Martina and Mendl, Christian B},
  journal={arXiv preprint arXiv:2312.08861},
  year={2023}
}

@book{williams1998explorations,
  title={Explorations in quantum computing},
  author={Williams, Colin P and Clearwater, Scott H and others},
  year={1998},
  publisher={Springer}
}

@article{rudolph2023decomposition,
  title={Decomposition of matrix product states into shallow quantum circuits},
  author={Rudolph, Manuel S and Chen, Jing and Miller, Jacob and Acharya, Atithi and Perdomo-Ortiz, Alejandro},
  journal={Quantum Science and Technology},
  volume={9},
  number={1},
  pages={015012},
  year={2023},
  publisher={IOP Publishing}
}

@article{batsheva2023protes,
  title={PROTES: Probabilistic Optimization with Tensor Sampling},
  author={Batsheva, Anastasia and Chertkov, Andrei and Ryzhakov, Gleb and Oseledets, Ivan},
  journal={arXiv preprint arXiv:2301.12162},
  year={2023}
}

@book{back1996evolutionary,
  title={Evolutionary algorithms in theory and practice: evolution strategies, evolutionary programming, genetic algorithms},
  author={Back, Thomas},
  year={1996},
  publisher={Oxford university press}
}

@article{TheorLimQGdec,
  title = {Minimal universal two-qubit controlled-NOT-based circuits},
  author = {Shende, Vivek V. and Markov, Igor L. and Bullock, Stephen S.},
  journal = {Phys. Rev. A},
  volume = {69},
  issue = {6},
  pages = {062321},
  numpages = {8},
  year = {2004},
  month = {Jun},
  publisher = {American Physical Society},
  doi = {10.1103/PhysRevA.69.062321},
  url = {https://link.aps.org/doi/10.1103/PhysRevA.69.062321}
}

@article{Rakyta2022approaching,
  doi = {10.22331/q-2022-05-11-710},
  url = {https://doi.org/10.22331/q-2022-05-11-710},
  title = {Approaching the theoretical limit in quantum gate decomposition},
  author = {Rakyta, P{\'{e}}ter and Zimbor{\'{a}}s, Zolt{\'{a}}n},
  journal = {{Quantum}},
  issn = {2521-327X},
  publisher = {{Verein zur F{\"{o}}rderung des Open Access Publizierens in den Quantenwissenschaften}},
  volume = {6},
  pages = {710},
  month = may,
  year = {2022}
}

@article{cleve2000introduction,
  title={An introduction to quantum complexity theory},
  author={Cleve, Richard},
  journal={Collected Papers on Quantum Computation and Quantum Information Theory},
  pages={103--127},
  year={2000}
}

@article{koike2010time,
  title={Time complexity and gate complexity},
  author={Koike, Tatsuhiko and Okudaira, Yosuke},
  journal={Physical Review A},
  volume={82},
  number={4},
  pages={042305},
  year={2010},
  publisher={APS}
}

@article{perelshtein2023nisq,
  title={{NISQ}-compatible approximate quantum algorithm for unconstrained and constrained discrete optimization},
 author={Perelshtein, M. R. and Pakhomchik, A. I. and Melnikov, Ar. A. and Podobrii, M. and Termanova, A. and Kreidich, I. and Nuriev, B. and Iudin, S. and Mansell, C. W. and Vinokur, V. M.},
  journal={Quantum},
  volume={7},
  pages={1186},
  year={2023},
  publisher={Verein zur F{\"o}rderung des Open Access Publizierens in den Quantenwissenschaften}
}

@article{perelshtein2020large,
  doi = {10.1002/andp.202200082},
  year = {2022},
  month = may,
  publisher = {Wiley},
  volume = {534},
  number = {7},
  pages = {2200082},
  author = {M. R. Perelshtein and A. I. Pakhomchik and A. A. Melnikov and A. A. Novikov and A. Glatz and G. S. Paraoanu and V. M. Vinokur and G. B. Lesovik},
  title = {Solving Large-Scale Linear Systems of Equations~by a Quantum Hybrid Algorithm},
  journal = {Annalen der Physik}
}

@article{white_paper_tq,
  title={Practical application-specific advantage through hybrid quantum computing},
  author={Perelshtein, Michael and Sagingalieva, Asel and Pinto, Karan and Shete, Vishal and Pakhomchik, Alexey and Melnikov, Artem and Neukart, Florian and Gesek, Georg and Melnikov, Alexey and Vinokur, Valerii},
  journal={arXiv preprint arXiv:2205.04858},
  year={2022}
}

@article{naumov2023tetra,
  title={Tetra-{AML}: automatic machine learning via tensor networks},
  author={Naumov, Alexey and Melnikov, Ar and Abronin, Vadim and Oxanichenko, Fedor and Izmailov, Kamil and Pflitsch, Markus and Melnikov, Alexey and Perelshtein, Michael},
  journal={arXiv preprint arXiv:2303.16214},
  year={2023}
}

@article{laskaris2023comparison,
  title={Comparison between Tensor Networks and Variational Quantum Classifier},
  author={Laskaris, Georgios and Melnikov, Artem A and Perelshtein, Michael R and Brasher, Reuben and Baeck, Thomas and Neukart, Florian},
  journal={arXiv preprint arXiv:2311.15663},
  year={2023}
}

@article{abronin2024tqcompressor,
  title={{TQC}ompressor: improving tensor decomposition methods in neural networks via permutations},
  author={Abronin, V and Naumov, A and Mazur, D and Bystrov, D and Tsarova, K and Melnikov, Ar and Oseledets, I and Dolgov, S and Brasher, R and Perelshtein, M},
  journal={arXiv preprint arXiv:2401.16367},
  year={2024}
}

@article{belokonev2023optimization,
  title={Optimization of chemical mixers design via tensor trains and quantum computing},
  author={Belokonev, Nikita and Melnikov, Artem and Podapaka, Maninadh and Pinto, Karan and Pflitsch, Markus and Perelshtein, Michael},
  journal={arXiv preprint arXiv:2304.12307},
  year={2023}
}

@article{morozov2023protein,
  title={Protein-protein docking using a tensor train black-box optimization method},
  author={Morozov, Dmitry and Melnikov, Artem and Shete, Vishal and Perelshtein, Michael},
  journal={arXiv preprint arXiv:2302.03410},
  year={2023}
}

@article{sagingalieva2023hybrid,
  title={Hybrid quantum {R}es{N}et for car classification and its hyperparameter optimization},
  author={Sagingalieva, Asel and Kordzanganeh, Mo and Kurkin, Andrii and Melnikov, Artem and Kuhmistrov, Daniil and Perelshtein, Michael and Melnikov, Alexey and Skolik, Andrea and Von Dollen, David},
  journal={Quantum Machine Intelligence},
  volume={5},
  pages={38},
  year={2023}
}

@article{termanova2024tensor,
  title={Tensor Quantum Programming},
  author={Termanova, A and Melnikov, Ar and Mamenchikov, E and Belokonev, N and Dolgov, S and Berezutskii, A and Ellerbrock, R and Mansell, C and Perelshtein, M},
  journal={arXiv preprint arXiv:2403.13486},
  year={2024}
}

@misc{Qiskit,
    author = {{Qiskit contributors}},
    title = {Qiskit: An Open-source Framework for Quantum Computing},
    year = {2023},
    doi = {10.5281/zenodo.2573505}
}

@InProceedings{pmlr-v151-anand-jain22a,
  title = 	 {{PAC Mode Estimation using PPR Martingale Confidence Sequences}},
  author =       {Anand Jain, Shubham and Shah, Rohan and Gupta, Sanit and Mehta, Denil and Nair, Inderjeet J. and Vora, Jian and Khyalia, Sushil and Das, Sourav and Ribeiro, Vinay J. and Kalyanakrishnan, Shivaram},
  booktitle = 	 {Proceedings of The 25th International Conference on Artificial Intelligence and Statistics},
  pages = 	 {5815--5852},
  year = 	 {2022},
  editor = 	 {Camps-Valls, Gustau and Ruiz, Francisco J. R. and Valera, Isabel},
  volume = 	 {151},
  series = 	 {Proceedings of Machine Learning Research},
  month = 	 {28--30 Mar},
  publisher =    {PMLR},
  url = 	 {https://proceedings.mlr.press/v151/anand-jain22a.html}
}

@article{berezutskii2025simulating,
  title={Simulating quantum circuits using the multi-scale entanglement renormalization ansatz},
  author={Berezutskii, Aleksandr V. and Luchnikov, Ilia A. and Fedorov, Aleksey K.},
  journal={Physical Review Research},
  volume={7},
  number={1},
  pages={013063},
  year={2025},
  doi={10.1103/PhysRevResearch.7.013063},
  publisher={American Physical Society}
}

@article{berezutskii2025tensor,
  title={Tensor networks for quantum computing},
  author={Berezutskii, Aleksandr and Liu, Minzhao and Acharya, Atithi and Ellerbrock, Roman and Gray, Johnnie and Haghshenas, Reza and He, Zichang and Khan, Abid and Kuzmin, Viacheslav and Lyakh, Dmitry and Lykov, Danylo and Mandr{\`a}, Salvatore and Mansell, Christopher and Melnikov, Alexey and Melnikov, Artem and Mironov, Vladimir and Morozov, Dmitry and Neukart, Florian and Nocera, Alberto and Perlin, Michael A. and Perelshtein, Michael and Steinberg, Matthew and Shaydulin, Ruslan and Villalonga, Benjamin and Pflitsch, Markus and Pistoia, Marco and Vinokur, Valerii and Alexeev, Yuri},
  journal={Nature Reviews Physics},
  volume={7},
  pages={581--593},
  year={2025},
  doi={10.1038/s42254-025-00853-1},
  publisher={Springer Nature}
}

@article{vidal2003efficient,
  title={Efficient classical simulation of slightly entangled quantum computations},
  author={Vidal, Guifr{\'e}},
  journal={Physical Review Letters},
  volume={91},
  number={14},
  pages={147902},
  year={2003},
  doi={10.1103/PhysRevLett.91.147902},
  publisher={American Physical Society}
}

@article{schollwock2011density,
  title={The density-matrix renormalization group in the age of matrix product states},
  author={Schollw{\"o}ck, Ulrich},
  journal={Annals of Physics},
  volume={326},
  number={1},
  pages={96--192},
  year={2011},
  doi={10.1016/j.aop.2010.09.012},
  publisher={Elsevier}
}

\clearpage
\appendix
\section{Success Probability}\label{a1}

Consider $\mathrm{H} = \sum_{k = 1}^{2^n}\limits f \left( x_{k} \right) \ketbra{k}$, the exact operator representing the discretized function values on its diagonals. We appropriately scale its elements such that, $\norm{\mathrm{H}} = 1$. In our approach, we obtain a unitary MPO approximation, $\mathrm{U}_{\mathrm{mpo}}$, by minimizing Eq.~\eqref{eq9}. Let $n_a$ represent the number of ancillary qubits required to cast $\mathrm{U}_{\mathrm{mpo}}$ into a quantum circuit. For simplifying the notation, we shall call this circuit $ \mathrm{U}$. The input state for this circuit is prepared as: $\rho_0 = \left( \ketbra{+}\right)^{\otimes n} \otimes \left( \ketbra{0}\right)^{\otimes n_a}$.
For post-selection, we define the projection $\Pi = \left(\mathds{1}^{\otimes n} \otimes \ketbra{0}^{\otimes n_a}\right)$.

Let us now consider a single quantum power iteration:
\begin{equation}\label{A:eq1}
    \rho_1 = \mathrm{U} \rho_0 \mathrm{U}^\dagger.
\end{equation}
The state after post-selection therefore becomes: 
\begin{equation}\label{A:eq2}
    \tilde{\rho}_{1} = \dfrac{1}{P_1} \Pi \rho_{1} \Pi,  
\end{equation}
where $P_1$ represents the post-selection probability given by: 
\begin{equation}
    \begin{split}\label{A:eq3}
        P_1 &= \mathrm{Tr}\left(\Pi \rho_1 \right),\\
            &= \mathrm{Tr}\left(\Pi \mathrm{U} \rho_0 \mathrm{U}^{\dagger} \right).
    \end{split}
\end{equation}

Since $\rho_0 = \Pi \rho_0 \Pi$,
\begin{equation}\label{A:eq4}
    \begin{split}
        P_1 &= \mathrm{Tr}\left(\Pi  \mathrm{U}  \Pi \rho_0 \Pi \mathrm{U}^{\dagger} \right), \\
            &= \mathrm{Tr}\left(\Pi  \mathrm{U}^\dagger  \Pi  \mathrm{U}  \Pi \rho_0 \right).
    \end{split}
\end{equation}
Note that if the approximation errors in obtaining $\mathrm{U}_{\mathrm{mpo}}$ are small, then since $\mathrm{H}$ is Hermitian,
\begin{equation}\label{A:eq5}
    \Pi  \mathrm{U}^{\dagger}  \Pi  \approx  \Pi  \mathrm{U}  \Pi\approx \mathrm{H} \otimes \ketbra{0}^{\otimes n_a}.
\end{equation}
Therefore, by substituting in Eq.~\eqref{A:eq4} and simplifying, we obtain: 
\begin{equation}\label{A:eq6}
    \begin{split}
        P_1 &= \mathrm{Tr} \left( \left( \mathrm{H}^2 \otimes \ketbra{0}^{\otimes n_a} \right) \rho_0 \right), \\ 
            &= \bra{+}^{\otimes n}  \mathrm{H}^2 \ket{+}^{\otimes n},\\
            &= \dfrac{1}{2^n} \mathrm{Tr}\left( \mathrm{H}^2 \right). 
    \end{split}
\end{equation}

Now, we consider the general case. After each iteration, we post-select and input the resultant state for the next iteration. Therefore, at $p^{\mathrm{th}}$ iteration, we obtain: 
\begin{equation}\label{A:eq7}
    \begin{split}
        \rho_p &= \mathrm{U} \tilde{\rho}_{p-1}  \mathrm{U}^{\dagger}, \\
                &= \dfrac{1}{P_{p-1}} \left( \mathrm{U} \Pi \rho_{p-1}  \Pi \mathrm{U}^{\dagger} \right).
    \end{split}
\end{equation}
The post-selection probability at this stage is given by: 
\begin{equation}\label{A:eq8}
    \begin{split}
        P_{p} &= \dfrac{1}{P_{p-1}} \mathrm{Tr}\left( \Pi \mathrm{U} \Pi \rho_{p-1}  \Pi \mathrm{U}^{\dagger} \right), \\
                &= \dfrac{1}{P_{p-1}} \mathrm{Tr}\left( \left( \mathrm{H}^2 \otimes \ketbra{0}^{\otimes n_a} \right) \rho_{p-1} \right). 
    \end{split}
\end{equation}
However, $\rho_{p-1}$ can further be expanded similar to in Eq.~\eqref{A:eq6} to obtain: 
\begin{equation}\label{A:eq9}
    P_{p} = \dfrac{1}{P_{p-1} \cdot P_{p-2}  } \mathrm{Tr}\left( \left( \mathrm{H}^4 \otimes \ketbra{0}^{\otimes n_a} \right) \rho_{p-2} \right). 
\end{equation}

Repeating this step $p-1$ times recursively, we obtain the expression: 
\begin{equation}\label{A:eq10}
    \begin{split}
        P_{p} &= \dfrac{1}{\prod_{j = 1}^{p-1} P_{j}} \mathrm{Tr}\left( \left( \mathrm{H}^{2\left( p-1\right)} \otimes \ketbra{0}^{\otimes n_a} \right) \rho_{1} \right),\\ 
                &= \dfrac{1}{\prod_{j = 1}^{p-1} P_{j}} \mathrm{Tr}\left( \left( \mathrm{H}^{2p} \otimes \ketbra{0}^{\otimes n_a} \right) \rho_{0} \right),\\
                &= \dfrac{1}{2^n \prod_{j = 1}^{p-1} P_{j}} \mathrm{Tr} \left( \mathrm{H}^{2p} \right).
        \end{split}
\end{equation}
Re-arranging the terms, we obtain: 
\begin{equation}\label{A:eq11}
    \prod_{j = 1}^{p} P_{j} = \dfrac{1}{2^n} \mathrm{Tr} \left( \mathrm{H}^{2p} \right). 
\end{equation}
Notice that the term on the left hand side of the equation represents the success probability of our method starting from the initial state $\rho_0$. 
We can further simplify the expression by noticing that: 
\begin{equation}\label{A:eq12}
    \mathrm{Tr} \left( \mathrm{H}^{2p} \right) = \sum_{k = 1}^{2^n} f^{2p}\left(x_{k} \right). 
\end{equation}
Recall the trapezoidal rule for definite integrals. For the uniform grid $x_k = a + \left(k-1\right)\left(b-a\right)/N$, $k = 1, \ldots, N$, which contains $a$ but excludes $b$:

\begin{equation}\label{A:eq13}
    \resizebox{.85\hsize}{!}{$
    \displaystyle \int_{a}^{b} g\left(x\right) dx = \dfrac{b-a}{N} \left[ \sum_{k = 1}^{N} g\left(x_{k} \right) + \dfrac{g\left(b\right) - g\left(a\right)}{2}\right] + \mathcal{O}\left(N^{-2}\right).
    $}
\end{equation}

Substituting and rearranging in Eq.~\eqref{A:eq11} with $N = 2^n$, we obtain the final expression:
\begin{equation}\label{A:eq14}
    \resizebox{.85\hsize}{!}{$
     \displaystyle P_{\mathrm{success}} \approx \dfrac{1}{b-a} \int_{a}^{b} f^{2p}\left(x\right) dx + \dfrac{f^{2p}\left(a\right) -  f^{2p}\left(b\right)}{2^{n+1}},
     $}
\end{equation}
where the omitted quadrature error is $\mathcal{O}\left(4^{-n}\right)$.
Notice that the expression becomes independent of $n$ up to an exponentially small boundary correction, allowing us to arbitrarily set the number of qubits for domain discretization. The analysis above assumes that the unitary MPO, $\mathrm{U}_{\mathrm{mpo}}$, approximates $\mathrm{H}$ accurately \emph{and} that the encoded block satisfies $\Pi \mathrm{U} \Pi \approx \mathrm{H} \otimes \ketbra{0}^{\otimes n_a}$ with $\norm{\mathrm{H}} = 1$. In practice, the variational minimization of Eq.~\eqref{eq9} yields a block $\tilde{\mathrm{U}} \approx \mathrm{H}/c^{\ast}$ with a scale factor $c^{\ast} \geq \norm{\mathrm{H}}$ (up to the fitting accuracy, since any block of a unitary has operator norm at most one). In this case, every factor of $\mathrm{H}$ above is replaced by $\mathrm{H}/c^{\ast}$, and the success probability acquires an additional factor $\left( \norm{\mathrm{H}} / c^{\ast}\right)^{2p}$; the values of $c^{\ast}$ realized in our experiments are reported in Table~\ref{tab:2}. This suppression is absent for the analytic sine construction, for which $c^{\ast} = \norm{\mathrm{H}} = 1$, and the measured success probabilities reproduce Eq.~\eqref{A:eq14} to within numerical precision.

\section{Inferring The Optimum}\label{a2}

In this section, we consider the number of measurements required to correctly identify the optimum of the function with sufficiently high probability. To do this, recalling Eq.~\eqref{eq1} and Eq.~\eqref{eq2}, we assume that there is a unique global optimum that is observed with probability, $p_{max} = \max_{k} p_k = p_{k^*}$. For convenience, we denote the next highest probability as $p_{next}$. 

By measuring the final state of our qubits, we are effectively sampling from an unknown probability distribution. Since the number of measurements that can be taken will always be finite, there is always the possibility that shot noise will cause the wrong basis state ($k \neq k^*$) to be identified as the global optimum. This problem can be tackled by selecting a so-called mistake probability, $\delta_{\mathrm{m}}$ (not to be confused with the relative gap $\delta$ of the main text), and then implementing an optimal stopping procedure that continues to take more measurements, only terminating once the probability of misidentifying the optimum is less than or equal to $\delta_{\mathrm{m}}$. Stated positively, the procedure only terminates once the probability of correctly identifying the optimum is at least $1 - \delta_{\mathrm{m}}$. 

Such a procedure is provided in Ref.~\cite{pmlr-v151-anand-jain22a}. Referred to as PPR-1v1, it requires fewer measurements to be performed than alternative approaches. According to theorem 9 of Ref.~\cite{pmlr-v151-anand-jain22a}, the number of measurements required by PPR-1v1 is at most 
\begin{multline}\label{B:eq1}
     M \leq \frac{194.07\, p_{max}}{(p_{max} - p_{next})^2} \\ \ln \left( \sqrt{\frac{79.68(K - 1)}{\delta_{\mathrm{m}}}} \frac{p_{max}}{(p_{max} - p_{next})} \right),
\end{multline}
where $K = 2^n$ is the number of discretization points.

It is essential that $p_{max}$ and $p_{next}$ in Eq.~\eqref{B:eq1} are probabilities of the \emph{normalized} post-selected state, a point which was not accounted for in the previous version of this analysis. After $p$ successful iterations the measured distribution is
\begin{equation}\label{B:eq2}
    p^{\left(p\right)}_{k} = \dfrac{f^{2p}\left(x_{k}\right)}{\sum_{j = 1}^{2^n} f^{2p}\left(x_{j}\right)},
\end{equation}
and its normalization is precisely the quantity computed in Appendix~\ref{a1}: by Eq.~\eqref{A:eq11}, $\sum_{j} f^{2p}(x_{j}) = 2^{n} \prod_{j=1}^{p} P_{j} = 2^{n} P_{\mathrm{success}}\left(p\right)$. Therefore
\begin{equation}\label{B:eq3}
    p_{max} = \dfrac{f_{(1)}^{2p}}{2^{n} P_{\mathrm{success}}\left(p\right)}, \qquad
    p_{max} - p_{next} = p_{max}\,\gamma_{p},
\end{equation}
where $f_{(1)}$ and $f_{(2)}$ are the two largest discretized values, $\delta = ( f_{(1)} - f_{(2)} )/f_{(1)}$ is the relative gap of the main text, and $\gamma_{p} = 1 - \left(1 - \delta\right)^{2p}$ is the gap amplified by $p$ power iterations. Substituting into Eq.~\eqref{B:eq1},
\begin{equation}\label{B:eq4}
    M \lesssim \dfrac{194.07}{p_{max}\,\gamma_{p}^{2}} \ln\left(\cdots\right)
      = \dfrac{194.07 \cdot 2^{n} P_{\mathrm{success}}\left(p\right)}{f_{(1)}^{2p}\, \gamma_{p}^{2}} \ln\left(\cdots\right).
\end{equation}
We have verified Eq.~\eqref{B:eq3} numerically on the sine benchmark, where it holds to machine precision.

To see how this scales, consider the (truncated) Taylor expansion of a differentiable function, $f(x)$, about its global maximum, which for convenience and without loss of generality, we take to be at $x = 0$:
\begin{equation}
    f(x) \approx f(0) + \frac{1}{2}f^{''}(0)x^2,
\end{equation}
where $f^{''}(0)<0$. Writing $h = \left(b-a\right)/2^{n}$ for the grid spacing, Laplace's method gives $\sum_{j} f^{2p}(x_{j}) \approx \left( f^{2p}\left(0\right)/h \right) \sqrt{\pi f(0) / \left( p \lvert f^{''}(0) \rvert \right)}$, so that
\begin{equation}
    p_{max} \approx h \sqrt{\dfrac{p \lvert f^{''}(0)\rvert}{\pi f(0)}}, \qquad
    \gamma_{p} \approx \dfrac{p \lvert f^{''}(0)\rvert h^{2}}{f(0)},
\end{equation}
the latter for $p \lvert f^{''}(0)\rvert h^{2} \ll f(0)$. Substituting into Eq.~\eqref{B:eq4}, the upper bound on the number of measurements scales, up to logarithmic factors, as
\begin{equation}\label{B:eq6}
    M \sim \dfrac{f(0)^{5/2}}{\lvert f^{''}(0)\rvert^{5/2}}\, p^{-5/2} h^{-5} \sim 2^{5n} p^{-5/2}.
\end{equation}
Two factors here are absent from an analysis that ignores the normalization in Eq.~\eqref{B:eq2}: one power of $h^{-1}$ and one of $p^{-1/2}$, both originating in $p_{max} \propto h \sqrt{p}$. We have confirmed Eq.~\eqref{B:eq6} numerically: evaluating Eq.~\eqref{B:eq1} directly on the sine benchmark, the bound grows by a factor of $\approx 1.2 \times 10^{3}$ per two additional qubits (asymptotically $2^{10} \approx 1.0\times10^{3}$) and decreases by a factor of $0.029$ when $p$ is increased fourfold (asymptotically $4^{-5/2} = 0.031$).

Equation~\eqref{B:eq6} shows that identifying the discretized optimum \emph{exactly} costs a number of measurements growing exponentially with the number of qubits. This is not an artifact of the method: refining the grid makes the two largest values of a smooth objective increasingly degenerate, with $\delta \sim h^{2}$, so distinguishing them necessarily becomes harder. It is, however, a statement about exact identification of the argmax rather than about optimization. For continuous optimization the neighboring grid points under the peak are equally good candidate optima, and the relevant figure of merit is the probability of returning a point within a prescribed tolerance $\varepsilon$ of the continuum optimizer,
\begin{equation}\label{B:eq7}
    \mathrm{Pr}\left( \lvert x_{k} - x^{\ast} \rvert \leq \varepsilon \right)
    = \!\!\!\!\sum_{k \,:\, \lvert x_{k} - x^{\ast}\rvert \leq \varepsilon}\!\!\!\! p^{\left(p\right)}_{k}
    \;\approx\; \mathrm{erf}\left( \varepsilon \sqrt{\dfrac{p \lvert f^{''}(0)\rvert}{f(0)}} \right),
\end{equation}
which is a Riemann sum and is therefore \emph{independent of $n$}, approaching unity as $p$ grows. On the sine benchmark this probability equals $0.6388$, $0.6367$, $0.6365$ and $0.6365$ for $n = 10, 14, 18, 22$ at $\varepsilon = 0.2$ and $p = 10$, and rises to $0.9955$ at $p = 100$, confirming the $n$-independence to four digits. In this practically relevant sense a number of measurements independent of the discretization resolution suffices; it is only the exact identification of the discretized argmax, as certified by the stopping rule of Eq.~\eqref{B:eq1}, that carries the exponential cost.

This analysis reveals a trade-off between the depth of the quantum circuit, which increases with $p$, and the number of measurements, which decreases with $p$. A more sophisticated analysis of this trade-off, possibly incorporating the effects of imperfect logic gates, is left for future work.

\end{document}